\documentclass[11pt,a4paper]{article}
\usepackage[a4paper,margin=2.5cm]{geometry}
\usepackage[T1]{fontenc}
\usepackage{lmodern}
\usepackage{amsmath,amsbsy,amssymb,mathtools}
\usepackage{mathrsfs}
\IfFileExists{physics.sty}{\usepackage{physics}}{%
  \newcommand{\dd}{\mathop{}\!\mathrm{d}}
  \newcommand{\dv}[3][]{\frac{\mathrm{d}^{##1}##2}{\mathrm{d}##3^{##1}}}

  \newcommand{\cross}{\times}
  
}
\usepackage{bm}
\usepackage{float}
\usepackage{wrapfig}
\usepackage{graphicx}
\usepackage{xcolor}
\usepackage{subcaption}
\usepackage[most]{tcolorbox}
\usepackage{tikz}
\usepackage{tikz-3dplot}
\usetikzlibrary{arrows.meta}
\usepackage{booktabs}
\usepackage{cancel}
\usepackage{enumitem}
\usepackage{empheq}
\usepackage[numbers]{natbib}
\usepackage[colorlinks=true,allcolors=blue]{hyperref}
 
\newcommand{\precc}{\mathrel{\prec\!\!\ast}}
\providecommand{\coloneq}{\coloneqq}

\begin{document}
\title{The Weyl Tensor for Gravitational-Wave Causal Sets:\\ Thickness of Geodesic Analogs and Geodesic Deviation}

\author{Nauman Ibrahim$^1$ and Luca Bombelli$^2$\\[4pt]
{\small $^{1,2}$Department of Physics and Astronomy, University of Mississippi, University, MS, USA}\\[2pt]
{\small E-mail: $^1$sibrahim@olemiss.edu, $^2$bombelli@olemiss.edu}}
\date{}

\maketitle

\begin{abstract}
\noindent Causal set theory proposes that spacetime is replaced by a partially ordered set (a directed acyclic
graph) at the smallest scales. This graph is postulated to produce all the features of continuous geometry
above some discreteness scale. We develop an approach to sprinkling causal sets into linearized plane
gravitational wave spacetimes, and report on the feasibility of using the analog of geodesic deviation to find
a component of the Weyl tensor for such causal sets. We take the collection of all longest chains between a
pair of elements to be the discrete analog of the timelike geodesic between them and find through numerical
simulations in Minkowski spacetime of various dimensions that the spread or `thickness' of the collection
increases with chain length. We estimate, assuming an ideal element-to-element spatial distance estimator and
a wave of amplitude $h_0 = 10^{-2}$, that $N_{\text{total}} \gtrsim 10^8$ elements are needed to overcome the
noise introduced by the thickness and detect the curvature. This is beyond the $\mathcal{O}(10^6)$ elements we
can currently simulate. Future work may show ways of extracting more information from the sprinkled causal set
and hence lowering the estimate for $N_{\text{total}}$.

\end{abstract}


\section{Introduction}\label{sec:intro}

\noindent Intuitively, the Weyl curvature of a Lorentzian manifold dictates how the three-dimensional shape of
a collection of test particles changes as they ``fall'' through spacetime, without changing their volume.
Showing how to estimate it for a causal set approximated by a restricted class of spacetimes, the plane
gravitational waves (GW), serves as the main motivation for the work in this paper. Our idea is to estimate
numerically the \(C_{x0x0}\) component of the Weyl tensor (other components can be expressed in terms of this
one due to the simplicity of the situation) by implementing geodesic deviation in such a causal set. This
means identifying two discrete geodesics in the causal set and estimating the distance between them along
their length---the second derivative of this distance should be proportional to \(C_{x0x0}\). 

Our motivation for finding an analog of the Weyl tensor stems from the belief that it will help strengthen a
restricted version of the \emph{Hauptvermutung}, which says that a causal set can embed into two different
manifolds \emph{only if} they have the same invariant quantities, $\mathcal{G}=\mathcal{G'}$,
associated with a given discrete invariant $\mathcal{O}$ in the causal set (cf. Ref.\ \cite{Surya:2019ndm},
p.\ 26). This makes sense because the same causal set should not be embeddable in two spacetimes with
different Ricci scalars, for example. But for one to claim this, one needs to have as many order invariants
corresponding to continuum invariants as possible.

Adapting the geodesic deviation method to causal sets, however, faces multiple obstacles. First, spatial
distance estimators for causal sets are statistical in nature and are thus subject to fluctuations. But before
one can begin characterizing these fluctuations for GW causal sets, one realizes that there is a bigger
problem at hand, that of the non-uniqueness of the discrete analogs of geodesics, longest maximal chains
(LMCs). Their length is known to be proportional to the proper time along the corresponding continuum
geodesic. However, there are many such chains and a priori there is no reason to pick one over another. If we
consider their collection to be the discrete analog of the geodesic, then their envelopes can be thought to
form a `tube' (see Sec.~\ref{sec:thickness}). Thus, if the tubes are too thick, then the fluctuations in the
deviation could be larger than the value of the deviation itself, so one must have a quantitative
characterization of the thickness to see under what circumstances that is not the case.

Working in intervals in Minkowski spacetime and treating the causal set as a directed acyclic graph, we
develop an algorithm to find all the elements along all the LMCs, and using the ensemble average of the mean
of their radial distances from the continuum geodesic, we define $\langle\delta\rangle$ to be the thickness of
the tube. We find that it scales as $H^{1-d\beta_d}$ where $H$ is the height of the interval and $d$ is the
dimension, and the exponent $\beta_d$ is found to be less than $1/d$. Interestingly, this tells us that
$\langle\delta\rangle$ grows with $H$, making one wonder, on a note separate from the question of geodesic
deviation, whether this trend continues to macroscopic scales, and how one is then to define geodesics in
causal sets. We do not address this question here but invite others to be stimulated by it.

Returning to the question of adapting the geodesic deviation method to GW causal sets, getting a quantitative
handle on the geodesic thickness is not enough. There are primarily two challenges. First, one has to develop
a deeper understanding of the continuum use of geodesic deviation. We find that using geodesic deviation to
characterize the curvature is suitable only in the ``long wavelength'' regime, where the separation, $L$, of
the geodesics is smaller than the wavelength, $\lambda$, of the wave. LIGO, with its kilometer-scale arms and
~100 Hz gravitational waves ($\lambda \sim 3000$ km), is in the long-wavelength regime, but LISA, where the
arm length is comparable to the wavelength, is not, and thus geodesic deviation is not used in its analysis. 

Second, the thickness of the geodesic tubes means that there are multiple elements at each `level' of each
tube, so measuring the separation means averaging over the many pairwise distances, and the challenge is to
model the fluctuations of this average. These fluctuations are present even if one assumes an ideal
element-to-element distance estimator. So, in order to assess the feasibility of the geodesic deviation
method, one must model them independently of the aforementioned fluctuations in element-to-element distance
estimates.

To model these fluctuations of the average, we carried out a statistical analysis of a general problem that we
call the ``two-spheres problem''---if one has two identical isotropic random distributions of points spread
around two points (centers) separated along the $x$-direction, and the centers themselves are not known but
one has access to pairwise distances between the points of the two spheres, how does one find the variance of
the average of the pairwise distances? This is discussed in Sec.~\ref{sec:feasibility} and
Appendix~\ref{app:error-propagation}. We find that the variance is given by $\text{Var}[\bar{d}] =
\left(\frac{1}{m}+\frac{1}{n}\right)\frac{\delta^2}{3} + \mathcal{O}(\delta^4/s^2)$, where $m$ and $n$ are the
numbers of points in the two spheres, $\delta$ is the mean distance of the points from their centers, and $s$
is the separation of the centers. This scales as $1/m$ for $m \sim n$, rather than as $1/(mn)$ as it would for
independent distances, because each point enters $n$ (or $m$) of the pairwise distances, which are therefore
correlated.

Having overcome these obstacles, one is in a position to find the number of points required in an ideal setup
for this fluctuation to be smaller than the actual acceleration in a proof-of-concept setup, where we define
the acceleration through a three-point finite difference numerical derivative. We find that this method
requires upwards of $\mathcal{O}(10^8)$ elements to overcome the noise of the average due to the thickness of
the tubes. 

Although we would have liked to find out otherwise, our result does not say that there is no other way of
constructing a discrete analog of the Weyl tensor; rather, it serves as motivation to continue looking. For
example, we can conceive of estimating the acceleration of the deviation by taking more measurements along the
tubes, i.e., using more information from the causal set than just a distance estimate at one level.

The rest of the paper is organized as follows. Section~\ref{sec:cst} reviews the essentials of causal set
theory. Section~\ref{sec:thickness} quantifies the thickness of the tubes of longest maximal chains in
Minkowski sprinklings. Section~\ref{sec:sprinkling} presents the sprinkling method for GW spacetimes.
Section~\ref{sec:gde} sets up geodesic deviation as a probe of the Weyl tensor and establishes its
limitations. Section~\ref{sec:feasibility} carries out the feasibility analysis for the proof-of-concept
setup, and Sec.~\ref{sec:discussion} closes with a discussion and outlook. The appendices collect the
derivation of the causal-relations formula and the error-propagation algebra.

\section{A brief review of causal set theory}\label{sec:cst}

\noindent Causal set theory (CST) is a candidate theory of quantum gravity whose central conjecture is that
the geometry and topology of a continuum spacetime above some scale can be captured by a \emph{countable} set
of elements with causal relations among them (a partially ordered set) and the number of elements corresponds
to spacetime volume \cite{bombelli_space_time_1987}. Despite introducing many novel features to the concept of
``spacetime'', the theory has its roots in general relativity (GR) and also hopes to reproduce GR's
predictions at large scales. The insight its foundation is built on is that the metric only carries
information about two things at each point of spacetime---the volume scale (conformal factor) and the
light-cone structure. This was proven in the 1970s by Hawking, King, McCarthy and Malament, and is known as
the HKMM theorem (cf. Ref.\ \cite{Surya:2019ndm}, p.\ 2).\footnote{If this is the reader's first exposure to
the causal set hypothesis, we point the interested reader to the excellent book \cite{surya_causal_2025}.}

In mathematical terms we say that a causal set, $\mathcal{C}$, is a partially ordered set that is locally
finite, where the partial order represented by $\preceq$ dictates the causal relation of the elements, i.e.,
if $i\preceq j$ then $j$ is to the causal future of $i$.\footnote{In most of the literature on causal sets a
strict partial order $\prec$ is used, where $i\prec j$ if $i\preceq j$ and $i\neq j$ \cite{Bombelli_2021}.} By
local finiteness, we mean that for any $p,q\in \mathcal{C}$ the cardinality of $I(p,q)=\{i\mid p\prec i \prec
q\}$ is finite; $I(p,q)$ is called the \emph{interval} between $p$ and $q$ and is the discrete analog of the
Alexandrov set defined for two points in a Lorentzian manifold as $A(x,y):= I^+(x) \cap I^-(y)$. Local
finiteness is the radical departure from the continuum, for there are uncountably many points in any interval
in the continuum; it is also what allows the number of points to be associated with volume. Similarly, we can
define a further relation $\precc$, called a link, where $i\precc j\implies i\prec j$ and $I(i,j)=\emptyset$
which denotes that $i$ and $j$ have no element between them. Such a relation does not exist in the continuum,
because, again, there are always an uncountable number of points between any pair.

The causal sets studied in this paper are generated from continuum spacetimes by \emph{sprinkling}. Sprinkling
points into a Lorentzian manifold is a manifestation of the Poisson point process (PPP). Generating points
through the PPP ensures that the resulting number of points in any given finite region follows a Poisson
distribution whose mean is proportional to the volume of the region. Sprinkling in practice means generating
coordinates of points of a manifold (in a fixed frame) and imposing a partial order on these points based on
their causal relation in the manifold, i.e., which point is in the future of which other points
\cite{Surya:2019ndm}. The coordinates are generated as random numbers distributed according to the manifold's
volume measure, $\sqrt{-\text{det }g_{\mu\nu}} \equiv \sqrt{-g}$, taken as a probability density function; if
the measure is not a function of some of the coordinates then those coordinates are generated uniformly. It
has been shown \cite{Surya:2019ndm} that the average density of points generated this way is
frame-independent. The other part of sprinkling, finding the causal relations, is not straightforward for a
general manifold, however. This part of sprinkling is discussed in Sec.~\ref{sec:sprinkling}.

The question of whether a causal set can indeed encode all the topological and geometrical information of a
Lorentzian manifold falls under the area of the kinematics of CST. An important part of research in kinematics
is constructing analogs of invariant continuum quantities like the scalar curvature and the action using
structures that are native to a causal set, e.g., the number of \emph{chains} of length $k$. These are ordered
elements in a subset of a causal set each related to the next; formally, a $k$-chain is a totally ordered set
of the form $\{p_1, p_2, \hdots, p_{k+1} \mid p_1\prec p_2\prec \hdots \prec p_{k+1}\}$ \cite{Aghili_2019}.
Apart from chains, intervals, which have been introduced above, can also be used to construct discrete
analogs. Thus, an important milestone for the community was constructing the discrete analog of the Ricci
scalar, $R(e)$, at an element $e$ in terms of the number of intervals of certain cardinalities it forms with
elements to its past. It then allowed for the construction of the Einstein--Hilbert action, which, as we know,
is the integral of $R$ (in the case of causal sets it would be a sum of $R$ over all elements). However, the
Ricci tensor, $R_{\alpha\beta}=g^{\nu\mu}R_{\mu\alpha\nu\beta}$, is only the trace part of the full curvature
tensor, the Riemann tensor, $R_{\mu\alpha\nu\beta}$. The non-trace part is called the Weyl tensor, and its
discrete analog, or that of any scalars constructed from it, has yet to be found.\footnote{In the process of
finalizing the draft of this paper, we became aware of an article by Nathan Madsen
\cite{madsen_reconstruction_2026}, which uses a different approach from ours, seemingly employing parallel
transport, to determine the full Riemann tensor for causal sets.}

\section{Discrete geodesics and the thickness of their tubes}\label{sec:thickness}

We begin with the question of how well the longest maximal chains (LMCs) between two elements of $\mathcal{C}$
approximate timelike geodesics between the corresponding points in the manifold $M$ into which $\mathcal{C}$
is sprinkled. In a causal set elements can either be related $(x\prec y)$ or linked $(x\precc y)$, where the
latter means that they are related and the interval $I(x,y) \coloneq \{z \mid x \prec z \prec y\}$ is empty,
i.e., they are ``direct neighbors''. Chains made exclusively of links are called maximal chains (or paths).
Discrete analogs of timelike geodesics are posited to be \emph{longest} maximal chains (LMCs) to match the
continuum idea that timelike geodesics are curves of maximum proper time; that is, if a link is to represent a
movement in time, discrete geodesics should have the largest number of links.

\begin{figure}[H]
    \centering
    \begin{subfigure}[b]{0.49\textwidth}
        \centering
        \begin{tikzpicture}[x=0.62cm,y=0.62cm,>={Stealth[length=5pt]}]
  \definecolor{lmcpurple}{RGB}{128,30,140}
  \definecolor{lev1}{RGB}{229,6,6}
  \definecolor{lev2}{RGB}{229,107,6}
  \definecolor{lev3}{RGB}{229,209,6}
  \definecolor{lev4}{RGB}{147,229,6}
  \definecolor{lev5}{RGB}{45,229,6}
  \definecolor{lev6}{RGB}{6,229,68}
  \definecolor{lev7}{RGB}{6,229,170}
  \definecolor{lev8}{RGB}{6,186,229}
  \definecolor{lev9}{RGB}{6,84,229}
  \definecolor{lev10}{RGB}{29,6,229}
  \definecolor{lev11}{RGB}{131,6,229}
  \draw[thick] (0,0) -- (4.50,4.50) -- (0,9.00) -- (-4.50,4.50) -- cycle;
  \draw[densely dashed, thin] (-0.651,-0.75) -- (-0.651,9.00);
  \draw[densely dashed, thin] (0.651,-0.75) -- (0.651,9.00);
  \draw[red, thick] (0,0) -- (0,9.00);
  \fill[black!40] (1.588,1.976) circle (1.6pt);
  \fill[black!40] (-2.106,2.672) circle (1.6pt);
  \fill[black!40] (2.630,2.940) circle (1.6pt);
  \fill[black!40] (-2.610,3.392) circle (1.6pt);
  \fill[black!40] (2.657,3.678) circle (1.6pt);
  \fill[black!40] (-2.961,3.704) circle (1.6pt);
  \fill[black!40] (3.473,3.789) circle (1.6pt);
  \fill[black!40] (1.837,4.072) circle (1.6pt);
  \fill[black!40] (3.947,4.243) circle (1.6pt);
  \fill[black!40] (2.018,4.329) circle (1.6pt);
  \fill[black!40] (2.458,4.375) circle (1.6pt);
  \fill[black!40] (-4.235,4.407) circle (1.6pt);
  \fill[black!40] (1.326,4.954) circle (1.6pt);
  \fill[black!40] (2.498,5.062) circle (1.6pt);
  \fill[black!40] (2.597,5.084) circle (1.6pt);
  \fill[black!40] (3.344,5.125) circle (1.6pt);
  \fill[black!40] (1.945,5.232) circle (1.6pt);
  \fill[black!40] (2.551,5.453) circle (1.6pt);
  \fill[black!40] (-3.286,5.468) circle (1.6pt);
  \fill[black!40] (1.030,5.473) circle (1.6pt);
  \fill[black!40] (3.017,5.679) circle (1.6pt);
  \fill[black!40] (2.694,5.703) circle (1.6pt);
  \fill[black!40] (-0.870,5.776) circle (1.6pt);
  \draw[black!28, thin] (-0.127,0.502) -- (0.350,1.615);
  \draw[black!28, thin] (-0.127,0.502) -- (0.176,1.667);
  \draw[black!28, thin] (-0.473,0.559) -- (0.176,1.667);
  \draw[black!28, thin] (-0.473,0.559) -- (-0.428,1.749);
  \draw[black!28, thin] (0.176,1.667) -- (0.224,2.049);
  \draw[black!28, thin] (-0.487,1.932) -- (-0.346,2.579);
  \draw[black!28, thin] (-0.487,1.932) -- (0.931,3.681);
  \draw[black!28, thin] (-0.346,2.579) -- (-1.864,4.153);
  \draw[black!28, thin] (-0.346,2.579) -- (0.676,4.569);
  \draw[black!28, thin] (-0.346,2.579) -- (-0.677,5.023);
  \draw[black!28, thin] (-1.323,3.303) -- (-0.677,5.023);
  \draw[black!28, thin] (-0.346,2.579) -- (-0.529,5.132);
  \draw[black!28, thin] (-1.323,3.303) -- (-0.529,5.132);
  \draw[black!28, thin] (-0.346,2.579) -- (-0.190,5.300);
  \draw[black!28, thin] (-1.323,3.303) -- (-0.190,5.300);
  \draw[black!28, thin] (0.931,3.681) -- (-0.190,5.300);
  \draw[black!28, thin] (-0.677,5.023) -- (-0.198,5.762);
  \draw[black!28, thin] (-0.529,5.132) -- (-0.198,5.762);
  \draw[black!28, thin] (-0.190,5.300) -- (-0.198,5.762);
  \draw[black!28, thin] (-1.327,6.065) -- (-0.119,7.279);
  \draw[black!28, thin] (-1.036,6.491) -- (0.327,7.937);
  \draw[black, line width=1.3pt] (0.000,0.000) -- (-0.127,0.502);
  \draw[black, line width=1.3pt] (0.000,0.000) -- (-0.473,0.559);
  \draw[black, line width=1.3pt] (-0.127,0.502) -- (-0.428,1.749);
  \draw[black, line width=1.3pt] (-0.473,0.559) -- (0.350,1.615);
  \draw[black, line width=1.3pt] (0.350,1.615) -- (0.224,2.049);
  \draw[black, line width=1.3pt] (-0.428,1.749) -- (-0.487,1.932);
  \draw[black, line width=1.3pt] (-0.487,1.932) -- (-1.323,3.303);
  \draw[black, line width=1.3pt] (0.224,2.049) -- (0.931,3.681);
  \draw[black, line width=1.3pt] (-1.323,3.303) -- (-1.864,4.153);
  \draw[black, line width=1.3pt] (0.931,3.681) -- (0.676,4.569);
  \draw[black, line width=1.3pt] (-1.864,4.153) -- (-1.876,5.153);
  \draw[black, line width=1.3pt] (0.676,4.569) -- (-0.198,5.762);
  \draw[black, line width=1.3pt] (-1.876,5.153) -- (-1.320,6.053);
  \draw[black, line width=1.3pt] (-0.198,5.762) -- (-0.296,5.880);
  \draw[black, line width=1.3pt] (-0.296,5.880) -- (-0.537,6.692);
  \draw[black, line width=1.3pt] (-1.320,6.053) -- (-1.327,6.065);
  \draw[black, line width=1.3pt] (-1.327,6.065) -- (-1.036,6.491);
  \draw[black, line width=1.3pt] (-1.036,6.491) -- (-1.139,7.119);
  \draw[black, line width=1.3pt] (-0.537,6.692) -- (-0.119,7.279);
  \draw[black, line width=1.3pt] (-1.139,7.119) -- (-0.194,8.310);
  \draw[black, line width=1.3pt] (-0.119,7.279) -- (0.327,7.937);
  \draw[black, line width=1.3pt] (0.327,7.937) -- (0.401,8.591);
  \draw[black, line width=1.3pt] (-0.194,8.310) -- (0.000,9.000);
  \draw[black, line width=1.3pt] (0.401,8.591) -- (0.000,9.000);
  \filldraw[lev1, draw=white, very thin] (-0.127,0.502) circle (2.8pt);
  \filldraw[lev1, draw=white, very thin] (-0.473,0.559) circle (2.8pt);
  \filldraw[lev2, draw=white, very thin] (0.350,1.615) circle (2.8pt);
  \filldraw[lev2, draw=white, very thin] (0.176,1.667) circle (2.8pt);
  \filldraw[lev2, draw=white, very thin] (-0.428,1.749) circle (2.8pt);
  \filldraw[lev3, draw=white, very thin] (-0.487,1.932) circle (2.8pt);
  \filldraw[lev3, draw=white, very thin] (0.224,2.049) circle (2.8pt);
  \filldraw[lev4, draw=white, very thin] (-0.346,2.579) circle (2.8pt);
  \filldraw[lev4, draw=white, very thin] (-1.323,3.303) circle (2.8pt);
  \filldraw[lev4, draw=white, very thin] (0.931,3.681) circle (2.8pt);
  \filldraw[lev5, draw=white, very thin] (-1.864,4.153) circle (2.8pt);
  \filldraw[lev5, draw=white, very thin] (0.676,4.569) circle (2.8pt);
  \filldraw[lev5, draw=white, very thin] (-0.677,5.023) circle (2.8pt);
  \filldraw[lev5, draw=white, very thin] (-0.529,5.132) circle (2.8pt);
  \filldraw[lev6, draw=white, very thin] (-1.876,5.153) circle (2.8pt);
  \filldraw[lev5, draw=white, very thin] (-0.190,5.300) circle (2.8pt);
  \filldraw[lev6, draw=white, very thin] (-0.198,5.762) circle (2.8pt);
  \filldraw[lev7, draw=white, very thin] (-0.296,5.880) circle (2.8pt);
  \filldraw[lev7, draw=white, very thin] (-1.320,6.053) circle (2.8pt);
  \filldraw[lev8, draw=white, very thin] (-1.327,6.065) circle (2.8pt);
  \filldraw[lev9, draw=white, very thin] (-1.036,6.491) circle (2.8pt);
  \filldraw[lev8, draw=white, very thin] (-0.537,6.692) circle (2.8pt);
  \filldraw[lev10, draw=white, very thin] (-1.139,7.119) circle (2.8pt);
  \filldraw[lev9, draw=white, very thin] (-0.119,7.279) circle (2.8pt);
  \filldraw[lev10, draw=white, very thin] (0.327,7.937) circle (2.8pt);
  \filldraw[lev11, draw=white, very thin] (-0.194,8.310) circle (2.8pt);
  \filldraw[lev11, draw=white, very thin] (0.401,8.591) circle (2.8pt);
  \fill (0,0) circle (3pt); \fill (0,9.00) circle (3pt);
  \node[above=3pt] at (0,9.00) {target};
  \draw[densely dashed, thin] (0,0) -- (0,-0.75);
  \draw[->] (0,-0.45) -- (0.651,-0.45) node[right] {$\delta$};
  \draw[->] (0,-0.45) -- (-0.651,-0.45) node[left] {$\delta$};
  \node[below] at (0,-0.85) {source};
  \draw[black!35, thin] (5.90,0) -- (5.90,9.00);
  \draw[black!35, thin] (5.70,0) -- (6.10,0) (5.70,9.00) -- (6.10,9.00);
  \node[font=\scriptsize, below] at (5.90,-0.05) {$t{=}0$}; \node[font=\scriptsize, above] at (5.90,9.05) {$t{=}H$};
  \path[fill=lev1, fill opacity=0.45, draw=lev1, line width=0.8pt] (5.73,0.461) rectangle (6.07,0.601);
  \path[fill=lev2, fill opacity=0.45, draw=lev2, line width=0.8pt] (5.73,1.612) rectangle (6.07,1.752);
  \path[fill=lev3, fill opacity=0.45, draw=lev3, line width=0.8pt] (5.73,1.920) rectangle (6.07,2.060);
  \path[fill=lev4, fill opacity=0.45, draw=lev4, line width=0.8pt] (5.73,2.579) rectangle (6.07,3.681);
  \path[fill=lev5, fill opacity=0.45, draw=lev5, line width=0.8pt] (5.73,4.153) rectangle (6.07,5.300);
  \path[fill=lev6, fill opacity=0.45, draw=lev6, line width=0.8pt] (5.73,5.153) rectangle (6.07,5.762);
  \path[fill=lev7, fill opacity=0.45, draw=lev7, line width=0.8pt] (5.73,5.880) rectangle (6.07,6.053);
  \path[fill=lev8, fill opacity=0.45, draw=lev8, line width=0.8pt] (5.73,6.065) rectangle (6.07,6.692);
  \path[fill=lev9, fill opacity=0.45, draw=lev9, line width=0.8pt] (5.73,6.491) rectangle (6.07,7.279);
  \path[fill=lev10, fill opacity=0.45, draw=lev10, line width=0.8pt] (5.73,7.119) rectangle (6.07,7.937);
  \path[fill=lev11, fill opacity=0.45, draw=lev11, line width=0.8pt] (5.73,8.310) rectangle (6.07,8.591);
  \node[lev1, font=\footnotesize, anchor=east] at (5.65,0.531) {1};
  \node[lev3, font=\footnotesize, anchor=east] at (5.65,1.990) {3};
  \node[lev5, font=\footnotesize, anchor=east] at (5.65,4.726) {5};
  \node[lev7, font=\footnotesize, anchor=east] at (5.65,5.967) {7};
  \node[lev9, font=\footnotesize, anchor=east] at (5.65,6.885) {9};
  \node[lev11, font=\footnotesize, anchor=east] at (5.65,8.450) {11};
  \node[lev2, font=\footnotesize, anchor=west] at (6.15,1.682) {2};
  \node[lev4, font=\footnotesize, anchor=west] at (6.15,3.130) {4};
  \node[lev6, font=\footnotesize, anchor=west] at (6.15,5.458) {6};
  \node[lev8, font=\footnotesize, anchor=west] at (6.15,6.378) {8};
  \node[lev10, font=\footnotesize, anchor=west] at (6.15,7.528) {10};
\end{tikzpicture}
        \caption{}
    \end{subfigure}\hfill
    \begin{subfigure}[b]{0.49\textwidth}
        \centering
        \begin{tikzpicture}[x=0.62cm,y=0.62cm,>={Stealth[length=5pt]}]
  \definecolor{lmcpurple}{RGB}{128,30,140}
  \draw[->] (-5.2,0) -- (5.7,0) node[below, font=\footnotesize] {space};
  \draw[->] (0,0) -- (0,10.40) node[right, font=\footnotesize] {time axis};
  \draw[thick] (0,0) -- (4.50,4.50) -- (0,9.00) -- (-4.50,4.50) -- cycle;
  \fill[black!40] (1.588,1.976) circle (2.0pt);
  \fill[black!40] (-2.106,2.672) circle (2.0pt);
  \fill[black!40] (2.630,2.940) circle (2.0pt);
  \fill[black!40] (-2.610,3.392) circle (2.0pt);
  \fill[black!40] (2.657,3.678) circle (2.0pt);
  \fill[black!40] (-2.961,3.704) circle (2.0pt);
  \fill[black!40] (3.473,3.789) circle (2.0pt);
  \fill[black!40] (1.837,4.072) circle (2.0pt);
  \fill[black!40] (3.947,4.243) circle (2.0pt);
  \fill[black!40] (2.018,4.329) circle (2.0pt);
  \fill[black!40] (2.458,4.375) circle (2.0pt);
  \fill[black!40] (-4.235,4.407) circle (2.0pt);
  \fill[black!40] (1.326,4.954) circle (2.0pt);
  \fill[black!40] (2.498,5.062) circle (2.0pt);
  \fill[black!40] (2.597,5.084) circle (2.0pt);
  \fill[black!40] (3.344,5.125) circle (2.0pt);
  \fill[black!40] (1.945,5.232) circle (2.0pt);
  \fill[black!40] (2.551,5.453) circle (2.0pt);
  \fill[black!40] (-3.286,5.468) circle (2.0pt);
  \fill[black!40] (1.030,5.473) circle (2.0pt);
  \fill[black!40] (3.017,5.679) circle (2.0pt);
  \fill[black!40] (2.694,5.703) circle (2.0pt);
  \fill[black!40] (-0.870,5.776) circle (2.0pt);
  \fill[lmcpurple] (-0.127,0.502) circle (2.0pt);
  \draw[lmcpurple,->,shorten >=2pt] (0,0.559) -- (-0.473,0.559);
  \fill[lmcpurple] (-0.473,0.559) circle (2.0pt);
  \draw[lmcpurple,->,shorten >=2pt] (0,1.615) -- (0.350,1.615);
  \fill[lmcpurple] (0.350,1.615) circle (2.0pt);
  \fill[lmcpurple] (0.176,1.667) circle (2.0pt);
  \draw[lmcpurple,->,shorten >=2pt] (0,1.749) -- (-0.428,1.749);
  \fill[lmcpurple] (-0.428,1.749) circle (2.0pt);
  \draw[lmcpurple,->,shorten >=2pt] (0,1.932) -- (-0.487,1.932);
  \fill[lmcpurple] (-0.487,1.932) circle (2.0pt);
  \node[anchor=east, font=\scriptsize, inner sep=1pt] at (-0.737,2.052) {$r_{6}$};
  \draw[lmcpurple,->,shorten >=2pt] (0,2.049) -- (0.224,2.049);
  \fill[lmcpurple] (0.224,2.049) circle (2.0pt);
  \draw[lmcpurple,->,shorten >=2pt] (0,2.579) -- (-0.346,2.579);
  \fill[lmcpurple] (-0.346,2.579) circle (2.0pt);
  \draw[lmcpurple,->,shorten >=2pt] (0,3.303) -- (-1.323,3.303);
  \fill[lmcpurple] (-1.323,3.303) circle (2.0pt);
  \draw[lmcpurple,->,shorten >=2pt] (0,3.681) -- (0.931,3.681);
  \fill[lmcpurple] (0.931,3.681) circle (2.0pt);
  \draw[lmcpurple,->,shorten >=2pt] (0,4.153) -- (-1.864,4.153);
  \fill[lmcpurple] (-1.864,4.153) circle (2.0pt);
  \node[anchor=east, font=\scriptsize, inner sep=1pt] at (-2.114,4.273) {$r_{11}$};
  \draw[lmcpurple,->,shorten >=2pt] (0,4.569) -- (0.676,4.569);
  \fill[lmcpurple] (0.676,4.569) circle (2.0pt);
  \draw[lmcpurple,->,shorten >=2pt] (0,5.023) -- (-0.677,5.023);
  \fill[lmcpurple] (-0.677,5.023) circle (2.0pt);
  \draw[lmcpurple,->,shorten >=2pt] (0,5.132) -- (-0.529,5.132);
  \fill[lmcpurple] (-0.529,5.132) circle (2.0pt);
  \draw[lmcpurple,->,shorten >=2pt] (0,5.153) -- (-1.876,5.153);
  \fill[lmcpurple] (-1.876,5.153) circle (2.0pt);
  \node[anchor=east, font=\scriptsize, inner sep=1pt] at (-2.126,5.273) {$r_{15}$};
  \fill[lmcpurple] (-0.190,5.300) circle (2.0pt);
  \fill[lmcpurple] (-0.198,5.762) circle (2.0pt);
  \draw[lmcpurple,->,shorten >=2pt] (0,5.880) -- (-0.296,5.880);
  \fill[lmcpurple] (-0.296,5.880) circle (2.0pt);
  \draw[lmcpurple,->,shorten >=2pt] (0,6.053) -- (-1.320,6.053);
  \fill[lmcpurple] (-1.320,6.053) circle (2.0pt);
  \draw[lmcpurple,->,shorten >=2pt] (0,6.065) -- (-1.327,6.065);
  \fill[lmcpurple] (-1.327,6.065) circle (2.0pt);
  \draw[lmcpurple,->,shorten >=2pt] (0,6.491) -- (-1.036,6.491);
  \fill[lmcpurple] (-1.036,6.491) circle (2.0pt);
  \draw[lmcpurple,->,shorten >=2pt] (0,6.692) -- (-0.537,6.692);
  \fill[lmcpurple] (-0.537,6.692) circle (2.0pt);
  \draw[lmcpurple,->,shorten >=2pt] (0,7.119) -- (-1.139,7.119);
  \fill[lmcpurple] (-1.139,7.119) circle (2.0pt);
  \fill[lmcpurple] (-0.119,7.279) circle (2.0pt);
  \draw[lmcpurple,->,shorten >=2pt] (0,7.937) -- (0.327,7.937);
  \fill[lmcpurple] (0.327,7.937) circle (2.0pt);
  \fill[lmcpurple] (-0.194,8.310) circle (2.0pt);
  \draw[lmcpurple,->,shorten >=2pt] (0,8.591) -- (0.401,8.591);
  \fill[lmcpurple] (0.401,8.591) circle (2.0pt);
  \fill (0,0) circle (2.6pt); \fill (0,9.00) circle (2.6pt);
  \node[below right, font=\footnotesize] at (0.05,-0.05) {$s$ (source)};
  \node[below left, font=\footnotesize] at (-0.05,-0.05) {$(0,0)$};
  \node[above right, font=\footnotesize] at (0.25,9.00) {$t$ (target)};
  \node[above left, font=\footnotesize] at (-0.25,9.00) {$(H,0)$};
  \node[anchor=west] at (-5.2,10.00) {\large $\mathbb{M}^2$};
  \node[anchor=north west, align=left, font=\scriptsize] at (-5.2,8.80) {Interval between $s$\\and $t$ with $N$\\elements inside};
  \fill[black!40] (2.5,2.0) circle (2.0pt); \node[anchor=west, align=left, font=\scriptsize] at (2.7,2.0) {Elements not\\along the LMCs};
  \fill[lmcpurple] (2.5,0.9) circle (2.0pt); \node[anchor=west, align=left, font=\scriptsize] at (2.7,0.9) {Elements\\along them};
\end{tikzpicture}
        \caption{}
    \end{subfigure}
    \caption[Tube of LMCs and interval setup]{\small A sprinkling of $N=50$ elements into a two-dimensional
    causal diamond of height $H$ (source at the bottom, target at the top) for illustrative purposes; (a) and
    (b) show the same sprinkling. (a)~All the elements and the links between them along all the longest
    maximal chains (LMCs), of length 12 against the continuum geodesic (red). Level $j$ consists of the
    elements that are $j$ links away from the source; each level has its own colour, the elements of a level
    are unrelated to each other, and the elements that lie on no LMC are the grey dots. Since the LMCs share
    elements, these can be combined in many ways: this sprinkling has 38 LMCs, and just two of these many
    paths are drawn with dark lines. On the common time axis at the right, each bar spans the $t$-coordinate
    range of one level, also showing overlaps in $t$ (here 5\,\&\,6, 8\,\&\,9 and 9\,\&\,10). The dashed lines
    are the boundaries of the tube at $\pm\delta$, where $\delta$ is the mean of the radial coordinates $r_i$
    of all the elements on the LMCs other than the source and the target (arrows in (b)). The boundaries are
    not intrinsic to the causal set. (b) The arrows show the radial coordinates $r_i$ of the elements along
    the LMCs, numbered by increasing $t$ (all 27 are drawn, only some are labelled). One can either consider
    all the elements or the middle third to include in the mean, giving two slightly different measures of
    thickness.}
    \label{fig:tube-setup}
\end{figure}
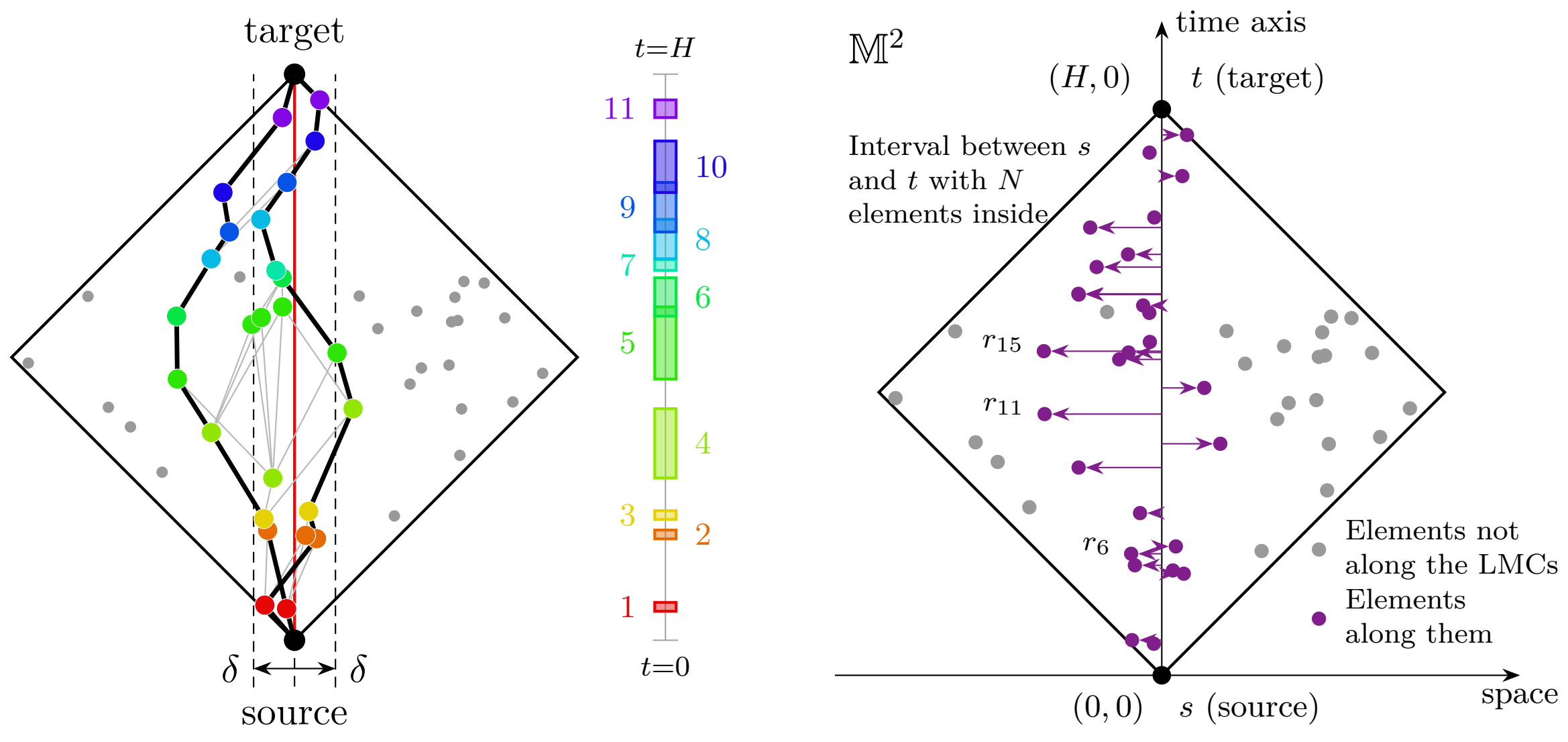

A key difference between the continuum geodesics and LMCs is that (as their name suggests) LMCs are not
unique: in general there are multiple longest maximal chains of the same length between two elements of
$\mathcal{C}$.\footnote{In fact their number grows exponentially with the number $N$ of elements in the
interval between the defining elements, as mentioned in Ref.\ \cite{rideoutwallden2009} and confirmed by us
through numerical work. The number of elements that make them up obviously grows much more slowly and is less
than $N$, thus the elements are shared between the paths.} Indeed, one can think of their collection as
forming a ``tube'' around the geodesic (Fig.~\ref{fig:tube-setup}(a)).

The question of the extent to which the correspondence between LMCs and timelike geodesics is a good one was
first rigorously addressed by Brightwell and Gregory (1991) \cite{BG} for the \emph{length} of the LMCs (i.e.,
the number of links they contain), $n_{C}$, in causal sets sprinkled into $\mathbb{M}^d$ ($d$-dimensional
Minkowski spacetime). They showed that for the longest paths in causal sets sprinkled into causal diamonds of
the same volume, $V$, the expectation value $\langle n_{C} \rangle$ is proportional to the height, $H$ (proper
time), of the diamond in the infinite density limit. Ilie, Thompson and Reid (2006) \cite{ilie_numerical_2006}
checked whether they could find a path that came arbitrarily close to the continuum geodesic in the infinite
density limit. Here, however, we investigate the \emph{profile} of \emph{all} the LMCs at fixed density, by
studying the \emph{radial distribution} of the elements that make up the LMCs, i.e., calling the envelope of
the LMCs a ``tube'', we find its ``thickness'' (Fig.~\ref{fig:tube-setup}(a)).

This is of interest because, for example---and this is how the question arose for us---one may want to measure
the separation, in the spirit of geodesic deviation, between two discrete geodesics in a causal set.
Specifically, could we estimate the appropriate components of the Weyl curvature in a causal set sprinkled
into a gravitational wave spacetime by finding the second numerical derivative of the separation? How would
one define the separation between two LMCs (tubes), each of which is by definition spread around the
corresponding continuum geodesic, without knowing their shape and thickness?

To simplify our analysis we will work with sprinklings in $\mathbb{M}^d$ (for $d\in \{2,3,4\}$) instead of a
curved spacetime. Moreover, for this analysis we need only consider causal diamonds (intervals) between the
elements that define the ends of the LMCs (`source', $s$, and `target', $t$) since any possible chain between
them must be inside the diamond, which is the whole region of shared future/past between them. We choose
coordinates such that the defining elements lie on the $t$-axis at $t=0$ and $t=H$ respectively (not to be
confused with the label `$t$' for the target element) (Fig.~\ref{fig:tube-setup}(b)). Finally, our analysis
will mostly be numerical.

\subsection{The thickness \texorpdfstring{$\langle\delta\rangle$}{<delta>} as a power law}

For us $\langle\delta\rangle$ is the primary quantity of interest which is defined as follows: take the set of
\emph{all elements} (excluding $s$ and $t$) \emph{that make up all LMCs} between two elements, $s$ and $t$, of
one sprinkling, $\mathcal{C}$, and find the mean of the radial coordinates of either (i) all the elements or
(ii) the middle third (based on their time coordinates), then $\langle\delta\rangle$ is the ensemble average
of that mean over many sprinklings. 

\begin{figure}[H]
    \begin{minipage}[c]{0.38\textwidth}
        \captionsetup{width=\linewidth}
        \caption[Plot of $\langle\delta\rangle / H$ vs.\ $N$ ]{Plot of $\langle\delta\rangle / H$ vs.\ $N$
        (all-element definition) showing the power-law trend for different values of $H$ (1 and 10) and
        confirming that $\langle\delta\rangle / H$ indeed only depends on the interval size $N$. Each point is
        an ensemble average over $\mathcal{O}(100)$ sprinklings.}
    \label{trend:a}
    \end{minipage}
    \hspace{0.5cm}
    \begin{minipage}[c]{0.57\textwidth}
        \includegraphics[width=\linewidth]{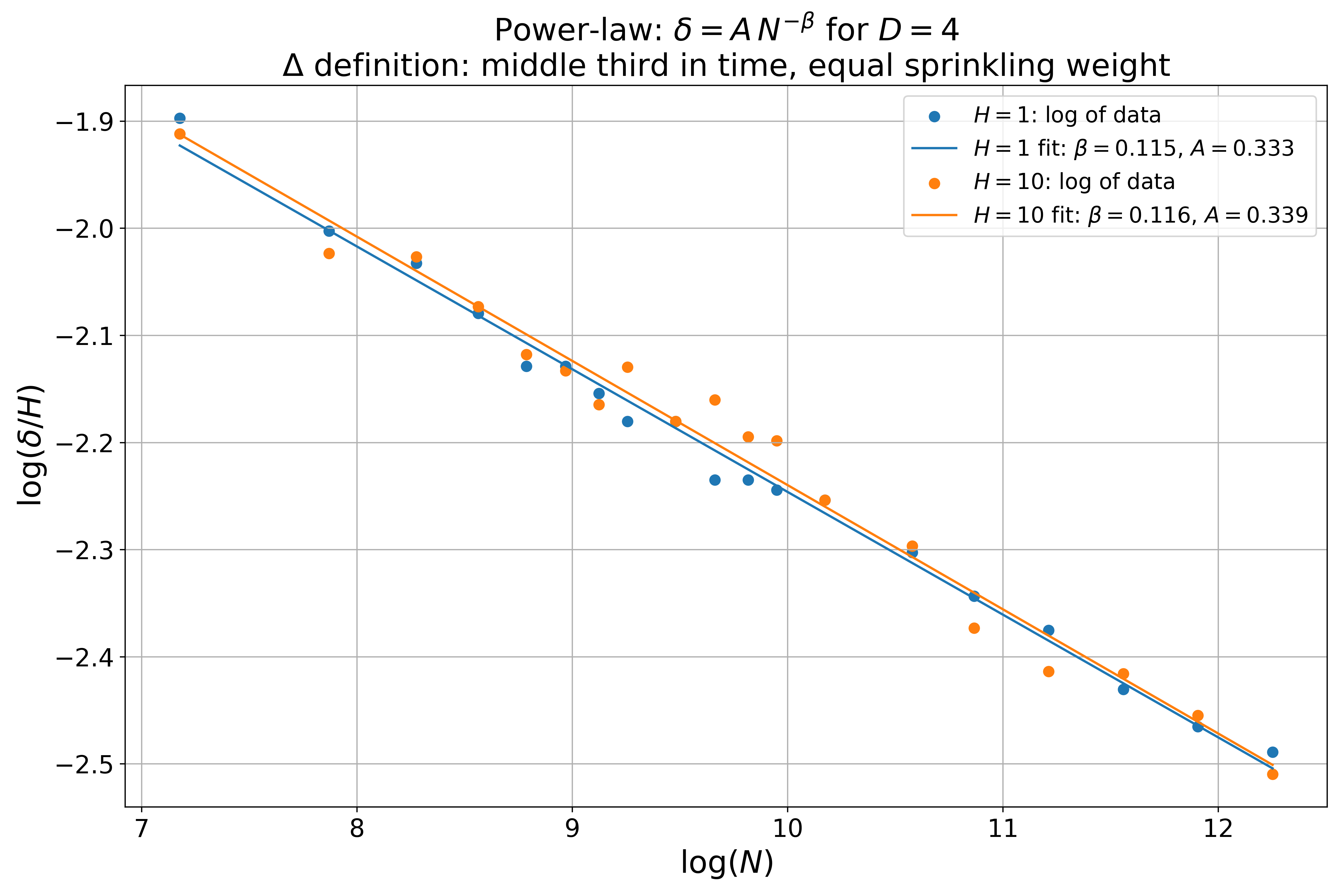}
    \end{minipage}
\end{figure}

\begin{figure}
    \centering
    \resizebox{0.6\linewidth}{!}{\begin{tikzpicture}[x=5.4cm,y=5.4cm,>={Stealth[length=5pt]}]
  \definecolor{lmcpurple}{RGB}{128,30,140}\definecolor{elemblue}{RGB}{0,120,255}\definecolor{levelgreen}{RGB}{0,110,80}
  \def\scaledinterval#1#2{%
    \begin{scope}[scale=#1]
      \draw[lmcpurple] (0,0) -- (0.5,0.5) -- (0,1) -- (-0.5,0.5) -- cycle;
      \foreach \x/\t in {-0.076/0.161, -0.169/0.264, 0.160/0.319, -0.056/0.326, 0.070/0.412, -0.211/0.419, 0.310/0.424, -0.063/0.454, 0.202/0.494, -0.322/0.536, 0.358/0.563, -0.151/0.622, 0.228/0.622, 0.110/0.692, -0.135/0.796, 0.053/0.895}
        \fill[lmcpurple] (\x,\t) circle (0.013);
      \fill[lmcpurple] (0,0) circle (0.013) (0,1) circle (0.013);
      \draw[red, thick, <->] (0,0) -- (0,1);
      \draw[elemblue, thick, {Bar[width=6pt]}-{Bar[width=6pt]}] (0,0.5) -- (0.135,0.5);
      \draw[->, thick] (0.015,0.78) to[bend left=18] (0.27,0.94) node[right] {$H_{#2}$};
      \draw[->, thick] (0.067,0.53) to[bend left=18] (0.42,0.73) node[right] {$\delta_{#2}$};
    \end{scope}}
  \begin{scope}[shift={(0,0)}] \scaledinterval{1}{B} \end{scope}
  \begin{scope}[shift={(1.25,0.2)}] \scaledinterval{0.6}{S} \end{scope}
  \node at (0.82,-0.04) {$\dfrac{\delta_B}{H_B} = \dfrac{\delta_S}{H_S}, \qquad N_B = N_S$};
  \draw[levelgreen, thick] (-0.62,-0.2) rectangle (1.85,1.12);
  \node[levelgreen, anchor=north west] at (-0.6,1.1) {$\mathbb{M}^2$};
\end{tikzpicture}}
    \caption[Scale invariance of intervals]{Two intervals, `small' and `big', with the same number of elements $N$ and identical set of relations, but different coordinate heights $H$.}
    \label{fig:scaling}
\end{figure}
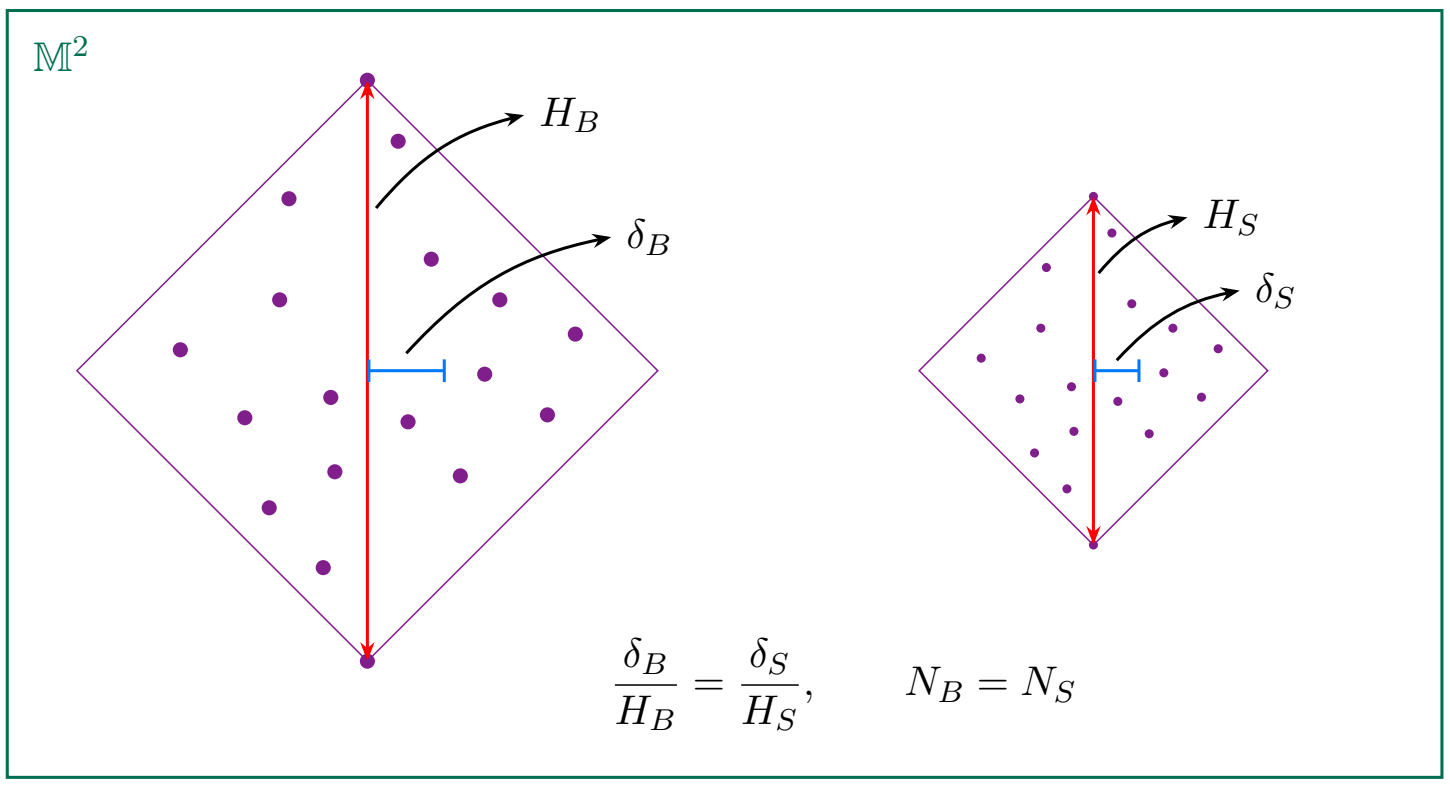

We find using numerical simulations that for either of these definitions (Fig.~\ref{trend:a}; the values
quoted are the mid-third ones, and both sets are in Table~\ref{tab:fit-params})
\begin{equation}
\frac{\langle\delta\rangle}{H} = A_d N^{-\beta_d} = A_d (\rho_d V_d)^{-\beta_d} \quad \text{with} \quad \beta_4 \approx 0.11, \; A_4 \approx 0.32,
\end{equation}
where $d$ is the dimension of the spacetime, $N$ is the number of elements sprinkled into the causal diamond
between $s$ and $t$,  $H$ is the proper time between them in the corresponding region of Minkowski
spacetime\footnote{From the numerical point of view, $H$ is the height of the sprinkling region from which the
causal sets have been generated.} (Fig.~\ref{fig:scaling}), $V_d = \zeta_d\,H^d$ is the volume of the diamond
(in four dimensions $\zeta_4 = \pi/24)$\footnote{More generally, $\zeta_d \equiv
S_{d-2}\frac{2^{1-d}}{d(d-1)}$, where $S_{d-2}$ is the volume of the unit $(d-2)$-sphere, $S_{d-2} \equiv
\frac{2\pi^{(d-1)/2}}{\Gamma((d-1)/2)}$ (see Ref.\ \cite{surya_causal_2025}, Ch.\,3).}, and $\rho_d=N/V_d$ is
the density of points. 

At first glance this result is as one would expect: as the density of points increases for a fixed volume
(consequently, height), i.e., in the continuum limit where $\rho\rightarrow \infty$,\footnote{An infinite
density would result in an infinite cardinality for the set. However, the points still remain
countable---unlike the true continuum where the points are uncountable.} the thickness of the discrete
geodesic goes to zero. However, one gets an unexpected result by casting this expression in terms of $H$ and
$\rho$:
\begin{equation}
    \boxed{\langle\delta\rangle = A'_d \cdot H^{1-d\beta_d},} \quad \text{where}\quad A'_d = A_d\,\zeta_d^{-\beta_d}\rho^{-\beta_d},
\end{equation}
\begin{align}
\text{which for $d=4$ is}\quad \langle\delta\rangle \approx A_4'\cdot H^{0.56}, \quad \text{with}\quad A_4' = A_4 \left(\frac{\pi}{24}\right)^{-0.11}\rho^{-0.11}, \nonumber
\end{align}
where $\langle\delta\rangle$ has dimensions of coordinate length. The unexpected aspect is the proportionality
to $H^{0.56}$ in 4D at a fixed density.\footnote{This density is expected to be one element per Planck volume
for the causal set that produces our spacetime.} It says that the thickness of the ``tube'' of LMCs for an
unaccelerated worldline increases unboundedly with the proper time as long as our spacetime can be
approximated by $\mathbb{M}^4$. In other dimensions, this result would hold as long as $\beta_d < 1/d$. In our
simulations for $d\in\{2,3,4\}$ we find $\beta_d$ to satisfy this bound robustly for alternative definitions
of $\delta$ as shown in the tables below. This is a very interesting result because if this is the case for
arbitrarily long chains, then the `tube'/region where the longest chains live becomes thicker and thicker as
the length of the chains increases, including becoming macroscopic. This has an implication for our
understanding of the geometry of causal sets and their relationship to continuum spacetimes, namely that
\textit{in order to identify longest chains as the discrete analog of the continuum geodesic, one must have a
condition other than the maximality of the chains.} We hope that this result will motivate further work on the
question of what the discrete analog of a geodesic is.

To estimate the uncertainty in $\beta_d$ we use the nonparametric bootstrap \cite{DavisonHinkley1997}: for
each $N$, we resample the sprinklings uniformly at random with replacement and recompute
$\langle\delta\rangle$, do this for all $N$ values and refit the power law; repeating this $B=2000$ times
gives $B$ values of $\beta_d$ and $A_d$, then the 95\% confidence interval is the range between the 2.5th and
97.5th percentiles of the $\beta_d$ values (Table~\ref{tab:beta-ci}).

\begin{table}[ht]
\centering
\small
\setlength{\tabcolsep}{7pt}
\renewcommand{\arraystretch}{1.2}
\begin{subtable}{0.48\textwidth}
\centering
\begin{tabular}{l|ccc}
\hline
 & $d=2$ & $d=3$ & $d=4$ \\
\hline
mid-third $A_{d}$  & 0.170 & 0.306 & 0.321 \\
all-elements $A_{d}$ & 0.136 & 0.254 & 0.287 \\
\hline
$\Delta A$ (all$-$mid) & $-0.034$ & $-0.052$ & $-0.034$ \\
\hline
\end{tabular}
\caption{$A_{d}$ comparison.}
\end{subtable}
\hfill
\begin{subtable}{0.48\textwidth}
\centering
\begin{tabular}{l|ccc}
\hline
 & $d=2$ & $d=3$ & $d=4$ \\
\hline
mid-third $\beta_{d}$  & 0.149 & 0.143 & 0.111 \\
all-elements $\beta_{d}$ & 0.152 & 0.146 & 0.119 \\
\hline
$\Delta\beta$ (all$-$mid) & $+0.003$ & $+0.003$ & $+0.008$ \\
\hline
\end{tabular}
\caption{$\beta_{d}$ comparison.}
\end{subtable}
\caption{Best-fit parameters $(A_{d},\beta_{d})$ in $\langle\delta\rangle/H = A_d\,N^{-\beta_d}$ ($H=1$),
compared across two definitions of $\delta$: ensemble average of the mean radial coordinate of elements in the
middle third of the chains by time coordinate versus ensemble average of the mean radial coordinate over all
the elements excluding $s$ and $t$. $\Delta$ is the difference in the parameters one gets from the two
definitions.}
\label{tab:fit-params}
\end{table}

\begin{table}[ht]
\centering
\small
\setlength{\tabcolsep}{7pt}
\renewcommand{\arraystretch}{1.2}
\begin{tabular}{l|ccc}
\hline
 & $d=2$ ($1/d=0.500$) & $d=3$ ($1/d=0.333$) & $d=4$ ($1/d=0.250$) \\
\hline
mid-third $\beta_d$ CI & $[0.137,\,0.161]$ & $[0.130,\,0.159]$ & $[0.105,\,0.118]$ \\
all-elements $\beta_d$ CI & $[0.144,\,0.162]$ & $[0.136,\,0.157]$ & $[0.114,\,0.124]$ \\
\hline
\end{tabular}
\caption{Bootstrap 95\% confidence intervals for $\beta_d$. The thickness grows if $\beta_d<1/d$ and we see that in all cases the entire interval lies strictly below $1/d$, confirming that $\langle\delta\rangle$ grows with chain length at fixed density.}
\label{tab:beta-ci}
\end{table}

The reason $\langle\delta\rangle/H$ depends only on $N$ is that, in general, any features of a uniformly
sprinkled causal set interval should only depend on the number of elements in the interval, $N$; a moment's
thought will convince the reader that this has to be the case. Then, for a fixed $N$, the coordinate height
$H$ of the interval becomes a mere scaling factor. This idea is depicted in Fig.\,\ref{fig:scaling}: the two
intervals, `small' and `big', are identical in terms of their causal structure and number of points, but have
been embedded in different-sized regions in $\mathbb{M}^2$; the figure argues that dimensionless features in
the two intervals must be the same, including for example, the ratio of $\delta$ to the coordinate height $H$.

A note on the definition of $\langle\delta\rangle$: the definitions we have used treat equally all sprinklings
with the same $N$, but one can demand more specificity and only treat sprinklings with the same chain length
as being equivalent. This would introduce an interesting feature that sprinklings with different $N$ but the
same chain length would have to be treated equally. There is also the question of the order of taking the mean
and the ensemble average, i.e., for sprinklings with the same chain length, whether the mean is first taken
over the elements of each sprinkling and then over the ensemble, or if it is taken across one level of all the
sprinklings and we find the average of those. These variations in the definition change the values of the
exponents and the prefactors slightly, though $\beta_d<1/d$ always holds. Ultimately, to decide between the
definitions one would have to see which one corresponds to the analytic calculation for the thickness one is
able to perform. One approach to the analytic calculation that comes from KPZ theory is discussed in the
following subsection.

\subsection{Shape of the tube and long-term behavior of the thickness}

\begin{figure}
    \begin{minipage}[c]{0.5\textwidth}
        \captionsetup{width=\linewidth}
    \caption[Plot of $r$ vs.\ $t$ in $d=4$]{Plot of $r$ vs.\ $t$ in $d=4$ of all the elements of $22$ sprinklings with the path length of sixty and $N=592k$. The angular coordinates of the elements are not plotted; each horizontal slice is the inside of $S^2$ with radius $\min(t,1-t)$. Roughly $1/10$th of the initial and the final points are close to the light cone. In the middle we see a bulge symmetric around $t=0.5$---much like an ellipsoid.}
    \label{shape}
    \end{minipage}
    \hfill
    \begin{minipage}[c]{0.36\textwidth}
        \includegraphics[width=\linewidth]{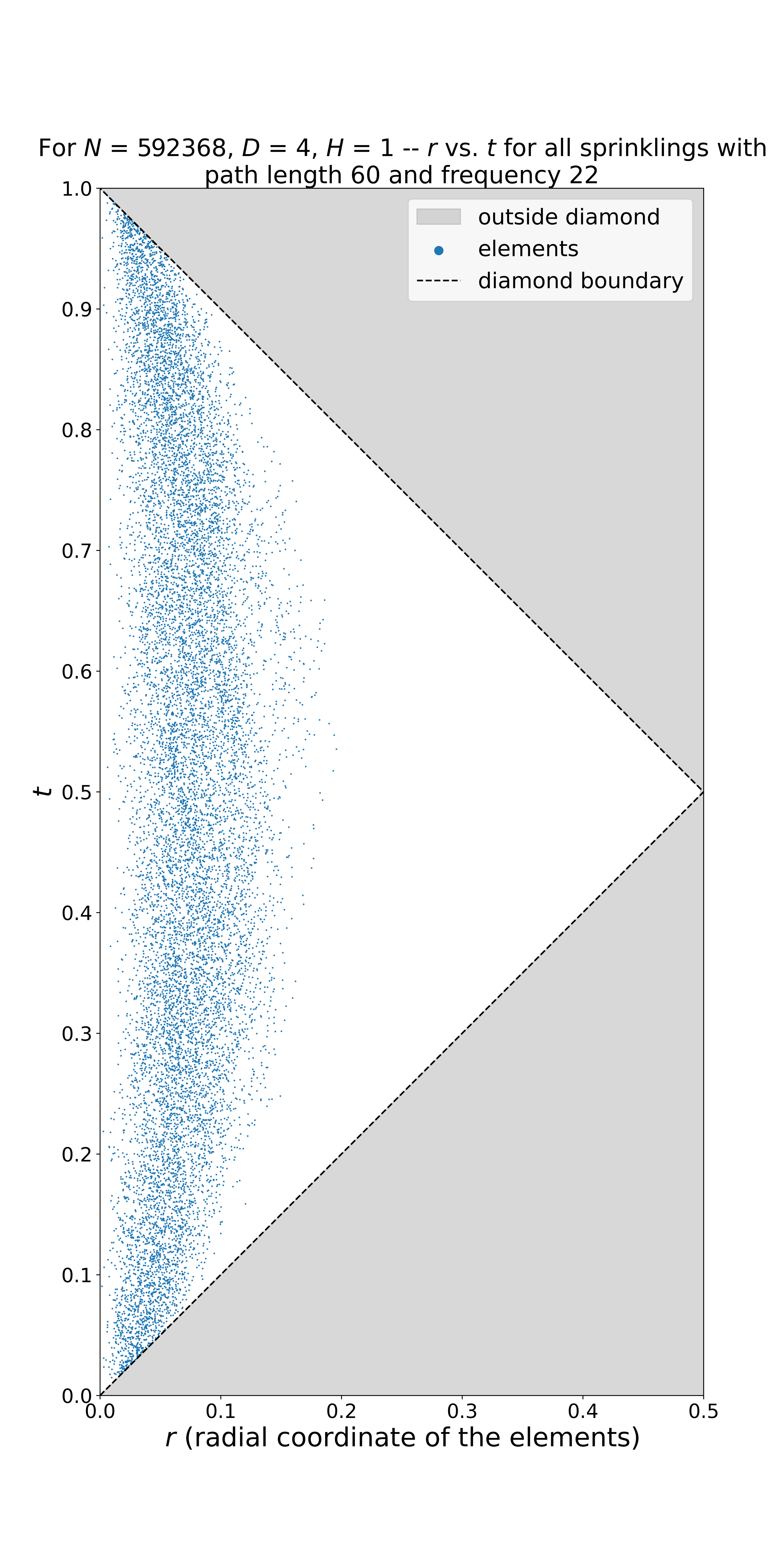}
    \end{minipage}
\end{figure}

We find that the shape of the tube is ``ellipsoidal'' in the $t$-$r$ plane, the middle being wide and ends
tapering off to zero (Fig.~\ref{shape}).

The question of whether this result holds for large intervals is still open, however. Our results come from
simulations; for example, in $d=4$ we have only been able to generate paths of length $k_{\max}\sim 80$ links.
The main limitation in how many elements we can simulate is the size of the causal set; the memory requirement
is of order $\mathcal{O}(N^2)$, which limits us to $N\sim\mathcal{O}(10^6)$. And according to Brightwell and
Gregory's result \cite{BG} the longest chain length, $k_{\textrm{max}}$, grows as $N^{1/d}$, so for higher
dimensions the chain length increases very slowly with $N$. So, to rigorously determine the long-term behavior
one would have to derive the behavior of the thickness analytically (see the note on this below). However,
there is a heuristic argument suggesting that thickness should keep on increasing for taller and taller
intervals.

The argument proceeds as follows. First, observe that if $a$ and $b$ are two elements on a longest path
between $s$ and $t$, then every longest path between $a$ and $b$ must itself be part of a longest path between
$s$ and $t$. This follows because any sub-segment of a longest path must itself be a longest path; otherwise,
a longer overall path could be constructed by replacing that sub-segment with a longer one.

Next, since longest paths are, on average, symmetric about the center of the interval, one should typically be
able to find two elements, one above and one below the midpoint of the interval by the same amount, and whose
spatial coordinates nearly coincide (picturing the causal set as embedded back into Minkowski space). If these
two elements are close enough in time, the thickness results of this section can be assumed to apply to them.
In that case, the longest chains connecting them are expected to bulge outward at their center. Since these
chains are themselves part of the longest chains between $s$ and $t$, the latter must bulge at the center as
well.

A note on deriving the power law analytically. We were in the process of deriving the power law analytically
by adapting a method for finding the mean number of paths of a given length at fixed density in an interval in
$\mathbb{M}^d$ \cite{Aghili_2019} when we were made aware, in a private communication from Dr.\ David Meyer,
that in $d=2$ in null coordinates the longest chains are equivalent to maximal up/right paths through a planar
Poisson point process. It is the classical and well-studied Hammersley process, or Poissonized Longest
Increasing Subsequence (LIS), or Poissonian Last Passage Percolation (LPP) setting, and the transversal
fluctuation exponent $\xi$ of that literature, which is our $1-d\beta_d$, is known to be $2/3$
\cite{hammond-sarkar-2020, johansson2000} up to a model dependent constant and polylogarithmic corrections.
The ``directed landscape'' is the scaling limit of a broad class of $(1+1)$-dimensional Kardar--Parisi--Zhang
(KPZ) models. Moreover, in higher dimensions it seems that the longest-chain problem is equivalent to the
zero-temperature directed polymer/last-passage percolation problem on a random medium with a built-in ``time
direction'' which strongly suggests that it is in the KPZ universality class for substrate dimension $d-1$,
specifically KPZ growth, and the various simulation results in the literature and other heuristics
\cite{oliveira-2022} allow one to estimate the exponents. These exponents seem to match our results.

\subsection{Identifying all LMC elements}

There are two challenges involved in determining how $\delta$ behaves. The first is to identify the elements
that lie along all the LMCs between a pair of elements $s \prec t$, and the second is to understand what
factors like $N,\,\rho,\,H$ it depends on and how. The latter was addressed for example through Fig.\
\ref{fig:scaling}; for the former, however, we take advantage of the fact that causal sets are directed
acyclic graphs (DAGs). Thus, to identify all the elements that lie on at least one of the LMCs between $s$ and
$t$, we develop a longest-path analog of the standard topological-order shortest-path algorithm for directed
acyclic graphs \cite[Sec.\ 24.2]{CLRS}, with the comparison reversed so that it maximizes the distance
\cite[p.\ 661]{sedgewick2011}. A short description of the algorithm is as follows. 

Using the fact that causal set elements can be labeled according to the order of their time coordinates, visit
the elements of the interval $[s,t]$ in this order. Assign to each element $x$ the number $d(x)$ of links in
the longest chain from $s$ to $x$, its `distance' from $s$, with $d(s)=0$. Every element in the past of $x$ is
visited before $x$, so each $d(y)$ with $y \prec x$ is already final when $x$ is reached; $d(x)$ is then
computed as
\begin{equation}
d(x) = 1 + \max_{\,s \preceq y \prec x} d(y).
\end{equation}
At each $x$ we also store \emph{all} the elements $y$ that attain the maximum distance. A single pass
therefore yields the length of the longest chains, $d(t)$, and tracing the stored predecessors back from $t$
recovers every element lying on at least one of the longest chains. This algorithm scales as
$\mathcal{O}(E+N)$, where $E\propto N^2$ is the number of edges, as opposed to a more straightforward
algorithm based on the powers of the adjacency matrix, which scales as $\mathcal{O}(N^3)$. We check our
algorithm against the \texttt{NetworkX} Python library's \texttt{all\_shortest\_paths} function with the
Bellman--Ford method with edges assigned negative weights and find our code to be faster and more
memory-efficient.

We reduce memory use by, for example, controlling individual bits within integers to store the relations matrix, hence
overcoming NumPy's restriction of using 8 bits to store a boolean type. The interested reader can access our
code at \href{https://github.com/sniafrmpk/cst}{\texttt{https://github.com/sniafrmpk/cst}}. We believe other
practitioners could benefit from the implementations we have developed. The repository also contains figures
of individual sprinklings in $d=2$ that show interesting features of the distribution of the LMC elements, for
example, the clustering of elements in places and their diverging into separate strands. An analysis of these
features might be of interest in its own right.

Another method of finding the elements along the longest chains is to first find the longest path length from
$s$ to all elements and from all elements to $t$ using some method that is fast, and then use the observation
that element $i$ of the causal set belongs to at least one LMC between $s$ and $t$ if and only if
\begin{equation}
    d(s,i) + d(i,t) = k_{\max},
\end{equation}
where $d(a,b)$ denotes the longest-chain length from $a$ to $b$ and $k_{\max} = d(s,t)$. We were made aware of
this criterion by Prof.\ Sorkin; it identifies all LMC elements without enumerating the chains explicitly.

\section{Sprinkling into a plus-polarized plane wave}\label{sec:sprinkling}

For a \emph{plus-polarized linear plane wave} (GW$_{+}$) the metric in transverse-traceless (TT) gauge
coordinates, $(t, x, y, z)$, representing a wave traveling in the $+z$ direction, in which freely falling test
particles sit at fixed $x, y, z$ is
\begin{equation*}
    \dd s^2 = -\dd t^2 + (1+h_{xx})\;\dd x^2+(1-h_{xx})\;\dd y^2 +\dd z^2 + \mathcal{O}(h_0^2),
\end{equation*}
where $h_{xx} = h_0\cos[\omega(t-z)]$ and $h_0$ is the amplitude of the wave. For typical wave detections, the
amplitude is $\sim 10^{-21}$, but in this work we will assume it is $\lesssim 10^{-2}$ because our concern is
to capture wave curvature in an idealized setting, not to relate it to experiment.

The same metric can be expressed in null coordinates, $u\coloneq (t-z)/\sqrt{2}$ and $v\coloneq
(t+z)/\sqrt{2}$, and in them $h_{xx}=h_{xx}(u)=h_0\cos(\omega' u)$ with $\omega'\coloneq\sqrt{2}\,\omega$, and
\begin{equation}\label{eq:2.1}
   g_{\mu\nu}= \begin{pmatrix}
    0&-1&0&0 \\
    -1&0&0&0 \\
    0&0&1+h_{xx}(u)&0 \\
    0&0&0&1-h_{xx}(u)
    \end{pmatrix}.
\end{equation}
The volume element is
\begin{equation}\label{eq:2.2}
    \sqrt{-g} =\sqrt{1-h_{xx}^2}=1-\frac{h_0^2}{2}\cos^2(\omega' u)+\mathcal{O}(h_0^4).
\end{equation}
Its departure from unity is of second order in $h_0$, the order already neglected in the metric, and is at
most $5\times10^{-5}$ for $h_0=10^{-2}$. To the order at which we work the coordinates of the sprinkled points
can therefore be generated uniformly in $(u,v,x,y)$.

\subsection{Finding causal relations via the geodesic tangent norm}

As mentioned earlier, generating coordinates for points is only half of the job of sprinkling into a
spacetime. The other half is finding the causal relations between the points, i.e., which points are in the
future light-cone of which others. Given only the metric and coordinates of two points in a general manifold
it is not easy to determine whether they are related. A relation would mean that there is a curve between the
points whose tangent vector is timelike or null everywhere along it.

One general way of finding such a curve between arbitrary points is to find the geodesic between them. Since
the norm of the tangent vector remains constant along a geodesic, our idea is to find the character of the
tangent vector at the initial point $p_1^\mu$ of the geodesic connecting the points, $p_1^\mu =
(u_1,v_1,x_1,y_1)$ and $p_2^\mu = (u_2,v_2,x_2,y_2)$, which we want to compare. If the norm is negative or
zero then the two points are related. Below we find this norm from quantities that are conserved along the
geodesic, without solving the geodesic equation in full.

For this method to work, these requirements need to be satisfied: (i) there is a geodesic between any two
points, (ii) it is unique. One way to ensure that they are satisfied is to sprinkle into a ``convex
neighborhood'': an open set $\mathscr{C}$ in a semi-Riemannian (Lorentzian) manifold such that ``for
\emph{any} two points $p, q$ of $\mathscr{C}$ there is a unique geodesic segment
$\sigma_{pq}\colon[0,1]\rightarrow M$ from $p$ to $q$ that lies entirely in $\mathscr{C}$''
\cite[p.~129]{ONeill-SemiR}. This gives (i) and (ii) immediately \cite[p.~146]{ONeill-SemiR}. The good news is
that each point of a Lorentzian manifold has a convex neighborhood (Ref.\ \cite{ONeill-SemiR}, p.\ 130). By a
``sufficiently small'' neighborhood in this work we mean one smaller than the length scale, $\mathit{l}_R$, at
which curvature manifests itself, which is usually $\sim1/\sqrt{R_{\mu\nu\alpha\beta}}$ where
$R_{\mu\nu\alpha\beta}$ is the largest component of the Riemann tensor. For the case of a plus-polarized
linearized gravitational wave (GW$_{+}$) all non-zero components of the Riemann tensor have a maximum absolute
value of $\frac12 \omega^2 h_0$, therefore $\mathit{l}_R \sim \frac{1}{\omega}\sqrt{2/h_0}$. Since $h_0$ is
small, $\mathit{l}_R$ is large.

If the metric does not depend on a coordinate $x^\sigma$, the covariant component
$W_\sigma=g_{\sigma\nu}W^\nu$ of the tangent vector $W^\mu=\dd w^\mu/\dd\lambda$ is constant along any
geodesic given by functions $w^\mu (\lambda) = (u( \lambda), v(\lambda), x(\lambda), y(\lambda) )$ \cite[Eq.\
3.167]{carroll_spacetime_2004}, whether it is timelike, null or spacelike.\footnote{We are using the
coordinate labels to identify the parametric functions of the geodesic which can cause some confusion but we
hope the reader can distinguish the two uses based on the context.} The metric (\ref{eq:2.1}) depends only on
$u$ (coordinate), so $W_v$, $W_x$ and $W_y$ are constant. Since $g_{uv}=-1$, the first of these is $W_v=-W^u$,
which means that $u(\lambda)$ changes at a constant rate along the geodesic. The other two are
$W_x=(1+h_{xx})W^x$ and $W_y=(1-h_{xx})W^y$, so the geodesic moves quickly in $x$ where $g_{xx}$ is small and
slowly where it is large, and the other way round in $y$. The norm $g_{\mu\nu}W^\mu W^\nu$ is a fourth
constant. The details of the calculation that follows are in Appendix~\ref{app:exact}.

The affine parameter is fixed only up to $\lambda\to m\lambda+b$ with constants $m$ and $b$, and we use this
freedom to make the geodesic leave $p_1$ at $\lambda=0$ and reach $p_2$ at $\lambda=1$. Any other choice
multiplies the norm by the positive constant $m^2$, which cannot change its sign. With this choice $W^u=\Delta
u=u_2-u_1$, and integrating $W^x=W_x/(1+h_{xx})$ and $W^y=W_y/(1-h_{xx})$ from $\lambda=0$ to $1$ gives $W_x$
and $W_y$ in terms of the coordinates of the two points. Finally, because the norm is constant it is equal to
its own integral over $\lambda$, and this integral can be carried out term by term. The result is
\begin{equation}\label{eq:geodesic-norm-exact}
    \boxed{g_{\mu\nu}W^\mu W^\nu=-2\Delta u\,\Delta v+\frac{(\Delta x)^2}{I_+}+\frac{(\Delta y)^2}{I_-}\,,}
    \qquad
    I_\pm\coloneq\frac{1}{\Delta u}\int_{u_1}^{u_2}\frac{\dd u}{1\pm h_0\cos(\omega' u)},
\end{equation}
\noindent where $\Delta v=v_2-v_1$, $\Delta x=x_2-x_1$, $\Delta y=y_2-y_1$ and $\omega'=\sqrt{2}\,\omega$. No
expansion in $h_0$ has been made, so Eq.\ \ref{eq:geodesic-norm-exact} is exact for the metric (\ref{eq:2.1}),
although the metric itself solves the vacuum Einstein equations only to first order in $h_0$. The integrals
$I_\pm$ have a closed form (Appendix~\ref{app:Ipm}), and for $\Delta u=0$ they take their limiting values
$1/(1\pm h_0\cos\omega'u_1)$. With this normalization, and because the norm of the tangent vector is constant
along a geodesic, $g_{\mu\nu}W^\mu W^\nu=\int_0^1 g_{\mu\nu}W^\mu W^\nu\,\dd\lambda=2\sigma(p_1,p_2)$, where
$\sigma$ is Synge's world function \cite{poisson_motion_2011}.

Eq.\ \ref{eq:geodesic-norm-exact} gives us a way of finding the causal relation between two points whose
coordinates are known, and thus a way of constructing the relations matrix (variously called the causal matrix
or the transitively complete adjacency matrix). The two integrals have a simple meaning: $1/I_+$ is the
harmonic mean of $g_{xx}=1+h_{xx}$ over the range of $u$ between the two points, and $1/I_-$ is the harmonic
mean of $g_{yy}$.

\section{Geodesic deviation as a probe of the Weyl tensor}\label{sec:gde}

The curvature of a spacetime manifold is fully characterized by the Riemann tensor. It can be decomposed into
a trace part, the Ricci tensor, and a trace-free part, the Weyl tensor. In four dimensions, the Riemann tensor
has twenty independent components, of which ten each belong to the Ricci and Weyl tensors. The Einstein field
equations algebraically relate the Ricci tensor to the matter stress-energy tensor; thus in vacuum spacetimes
the Ricci tensor vanishes, the Riemann tensor equals the Weyl tensor, and the geodesic deviation equation
(given below), which is in terms of the Riemann tensor, can be written in terms of the Weyl tensor. The
geometrical interpretation of the Weyl tensor is that it governs the evolution of the shear tensor for a
congruence of geodesics in the spacetime; in other words it determines how the spatial shape changes without
changing the volume (for more on its interpretation, see Lehmkuhl et al.\ \cite{Lehmkuhl_2024}). Below we give
the components of the Weyl tensor, $C^+_{\alpha\beta\mu\nu}$,  in the TT-gauge for a plus-polarized
gravitational wave moving in the $z$-direction, and the geodesic deviation equation, which relates the
acceleration of the deviation vector $\xi^\alpha$ between two neighboring geodesics to the curvature.

\begin{equation}
    C^{\textrm{+}}_{x0x0}=-C^{\textrm{+}}_{y0y0}=-C^{\textrm{+}}_{x0xz}=C^{\textrm{+}}_{y0yz}=C^{\textrm{+}}_{xzxz}=C^{\textrm{+}}_{yzyz}=\frac12\omega^2h_0\cos{\omega(t-z)},
\end{equation}
\begin{equation}
    \frac{\textrm{D}^2\xi^\alpha}{\dd \tau^2}\coloneq (\xi^\alpha{}_{;\mu}u^\mu)_{;\nu}u^\nu=-R^\alpha{}_{\lambda\mu\nu}u^\lambda \xi^\mu u^\nu.
\end{equation}
As an example, if two test particles are initially separated by $\epsilon$ in the $x$-direction in a
plus-polarized plane gravitational wave (GW$_{+}$), then we can set up coordinates in which, ignoring first-
and higher-order corrections to $u^\alpha$ in $h_0$ ($u_\alpha$ is being multiplied by the Riemann tensor
which is already first order in $h_0$), $\xi^\alpha=(0,\epsilon,0,0)$ and the tangent to the geodesics can be
taken to be $u^\alpha=(1,0,0,0)$, i.e., 
\begin{equation}
    \dv[2]{\xi^x}{\tau}\to\dv[2]{\xi^x}{t}, \quad R^\alpha{}_{\lambda\mu\nu}u^\lambda \xi^\mu u^\nu \to \,\epsilon R^x{}_{00x}=-\epsilon C^{\textrm{+}}_{x0x0}. \nonumber
\end{equation}
So, to first order in $h_0$ the deviation equation in the TT gauge reduces to \cite[p.\ 208]{Schutz:1985jx}
\begin{align}
   \dv[2]{\xi^x}{t} = -\frac12\epsilon\,\omega^2h_0\cos{\omega(t-z)}.
\end{align}
Thus, by estimating the separation $\xi$ at multiple times, we should be able to estimate numerically the
acceleration of the separation, and hence the Weyl tensor. 

\subsection{Limitations of the method}\label{sec:limitations}

First limitation: we work in the long-wavelength regime. It turns out that the geodesic deviation equation can
have higher-order corrections comprising the derivatives of the Riemann tensor multiplied with powers of the
separation vector $\xi$, such that for the gravitational wave case each successive term is a higher-order
power of (roughly)\footnote{From the exact expression for the higher-order geodesic deviation equation given
below one can see that while the Riemann tensor terms are exactly proportional to $1/\lambda^2$ and each
derivative multiplies by an additional factor of $1/\lambda$, the association of $L$ with the deviation vector
$\xi$ is a coordinate dependent one. However, for a linearized gravitational wave, this dependence would be of
the order of $h_0$ and so $\xi/\lambda \sim L/\lambda$.} $L/\lambda$---the separation of the geodesics over
the wavelength of the gravitational wave. For reference, the second-order geodesic deviation equation taken
from Ref.\ \cite{vines_geodesic_2015} (Vines) is:
    \begin{equation}
        \ddot{\xi}^\alpha + R^\alpha{}_{\beta\gamma\delta}u^\beta \xi^\gamma u^\delta + R_{\beta\gamma}{}^{\alpha}{}_{(\rho;\sigma)}\xi^\beta u^\gamma \xi^\rho u^\sigma + 2R^\alpha{}_{\beta\gamma\delta}\dot{\xi}^\beta\xi^\gamma u^\delta= \mathcal{O}(\xi,\dot{\xi})^3.
    \end{equation}
So, if $L \to \lambda$ the higher-order correction terms become increasingly important, so much so that for $L
\ge \lambda$ the whole series expansion breaks down. From this one can see that the framework of geodesic
deviation is only valid in the regime where $L \ll \lambda$, i.e., the long-wavelength regime.\footnote{This
is reflected in the fact that geodesic deviation is \emph{not} used to model the wave strain for LISA, the
planned space-based GW laser interferometer, whose arm lengths are many millions of kilometers and are of the
order of the wavelength of the waves it intends to detect. Refs.~\cite{baskaran_components_2004,
rakhmanov_response_2005, rakhmanov_fermi-normal_2014, virtuoso_beyond_2025, koop_gravitational_2014} discuss
the `long-wavelength' regime's limitations. See Refs. \cite{vines_geodesic_2015, kim_geodesic_2025} for more
on the higher-order geodesic deviation equation.} Within this regime, we choose to work only with the
leading-order term, i.e., $R^\alpha{}_{\beta\gamma\delta}u^\beta \xi^\gamma u^\delta$ so that there is a
linear relationship between the acceleration of the separation vector and the curvature, which is easy to
invert. The long-wavelength regime also places a limit on how large the initial separation, $\xi$, can be and
thus a limit on how large the acceleration can be made by choosing far-apart geodesics.

Second limitation: this proposal relies on an ideal element-to-element distance estimator. The only spatial
distance construction that addresses curvature, tested so far only for extrinsically curved hypersurfaces in
Minkowski spacetime, is due to Eichhorn et al. (2019) \cite{eichhorn_induced_2019}. It converts the minimal
light-cone volume suspended above two elements of an inextendible antichain (a collection of unrelated
elements satisfying certain conditions meant to represent a spatial hypersurface) into their spatial distance.
But this estimate, or any other one (two of which are mentioned in the concluding section), is expected to
have statistical fluctuations, which could wash away the signal of the wave. So in our analysis we assume an
ideal element-to-element distance estimator to isolate the noise from just the thickness of the geodesic
analogs.

Third limitation: realistically, we can only work with causal sets of 3--4 million points because they take up
to 1\,TB of RAM, which is the typical size for a node on a supercomputer. Using multiple nodes to run the code
is more challenging, and the gain in the number of points is small since the memory requirement scales as
$N^2$.

\section{Feasibility: the minimum number of elements required to detect the wave}\label{sec:feasibility}

Given the constraint that we must work in the $L \ll \lambda_{\text{GW}}$ regime, and the fact that the
thickness of the geodesic tubes depends on the density of points (which depends on $N_{\text{total}}$ and the
dimensions of the sprinkled region) and the length of the longest chains, we set out to investigate whether
the acceleration of the separation between the tubes can be captured in a causal set with a reasonable number
of points through a \emph{proof-of-concept setup} and its analysis.

The first step is to recognize that we can save on the number of points by working in a 4D ``album''-shaped
region whose length in two directions perpendicular to the separation between the geodesics is a few times the
thickness of the tubes, because the formation of the tubes is not affected by the presence of points outside
of them.

For simplicity, we will assume that the distances between the tops and bottoms of the tubes are the same and
are known without error.\footnote{The probability of finding geodesics that are exactly parallel at the crests
and troughs in a realistic causal set is negligible since this is equivalent to sprinkling two pairs of points
that lie on the tops and bottoms of these geodesics, which has negligible probability of happening in a
Poisson point process; but as we have said, this calculation is in the spirit of a proof-of-concept, where the
goal is to see, given an ideal setup, how robust the method we propose is.} The former ensures that the
geodesics are momentarily parallel when the phase, $\omega(t-z)$, is a multiple of $\pi$, i.e., at the crests
and troughs of the wave, and the latter helps reduce the sources of uncertainty. We will be interested in
checking if \textit{the error in the numerical second derivative of the separation at the middle due to the
thickness of the tubes can be made smaller than the value of the second derivative itself.}

\begin{figure}[h]
    \centering
    \includegraphics[width=\textwidth]{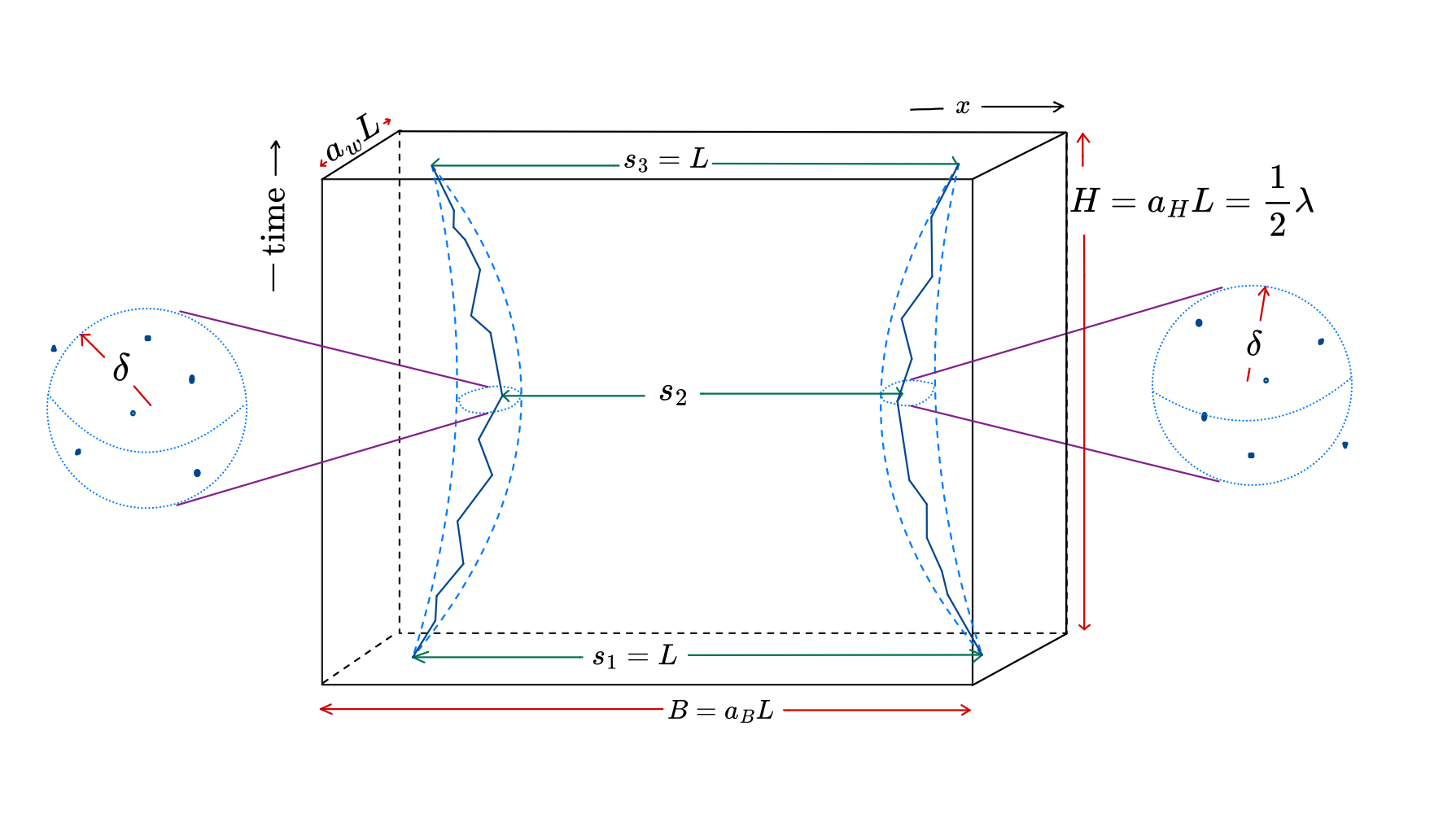}
    \caption[Our setup's diagram -- the `album' with spheres]{An illustration of the setup of the sprinkling
    region and the various dimensions. }
    \label{fig:album-with-spheres}
\end{figure}

\begin{figure}[htbp]
\centering
\includegraphics[width=0.8\textwidth]{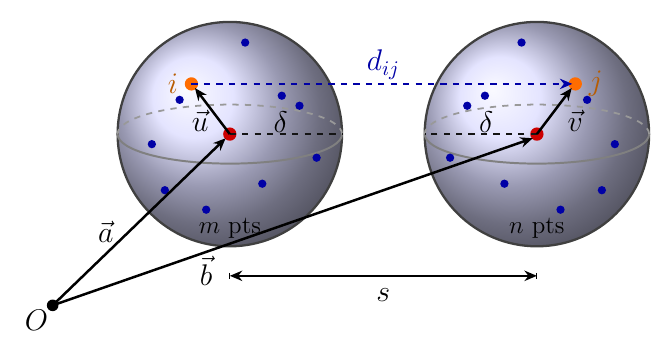}
\caption[Two spheres]{For a given sprinkling, at the levels corresponding to $t_2$ on both tubes, the elements
are isotropically distributed around the locations of the continuum geodesics' intersection with the $t = t_2$
hypersurface. The black outline of the spheres in the figure is not part of the causal set or the continuum
geometry; it represents the mean distance $\delta$ of the elements from the location of the continuum
geodesics. It is unlikely for an element to be exactly at the center of the spheres but we find that the
distance between the centers $s_2$ is approximately the mean of the pairwise distances of all the elements,
$\bar{d}$. We are interested in the variance of the mean for repeated samplings because that will tell us how
reliable our estimate of $s_2$ is for a given sprinkling.}
\label{fig:spheres}
\end{figure}

We want to estimate $C_{x0x0}$ at $t_2$ using the equation $\ddot{s}/s = -C_{x0x0}$. This is a gauge-invariant
equation in linearized gravity (for the reasons, see Ref.\ \cite[p.\ 208]{schutz_first_2022}). However, when
we measure $\ddot{s}/s$ at some `level' corresponding to time $t_2$ its value will have a spread because there
will be multiple element-to-element distance estimates for $s_2$ between the various elements at the `levels'
corresponding to $t_2$ on both the tubes. So, to be confident about a detection in a given sprinkling we would
like the spread $\sigma(\ddot{s}_2/s_2)$ to be less than the value of $\ddot{s}_2/s_2$ itself. Thus our
\textbf{detection condition} is
\begin{equation}\label{eq:detection_condition}
    \boxed{\sigma\!\left(\ddot{s}_2/s_2\right) < \left|\ddot{s}_2/s_2\right|.}
\end{equation}
Precisely, $\ddot{s}_2$ is defined via the three-point midpoint formula for the second derivative,
\begin{equation}
    \ddot{s}_2 = \frac{s_1 - 2s_2 + s_3}{(\Delta t)^2}, \qquad \Delta t = \frac{H}{2} = \frac{\lambda}{4},
\end{equation}
where $s_1 = s_3 = L$ are the separations at the bottom and top of the tubes. We choose $H = \lambda/2$, and
not more, to minimize the volume of the sprinkling region while still getting sufficient change in separation,
and choose the three-point formula so that there is only one source of error, the estimate of $s_2$. We define
$s_2$ as the mean of the element-to-element distances,
\begin{equation}
    s_2 \coloneq \bar{d} = \frac{1}{mn}\,\sum_i^m \sum_j^n d_{ij},
\end{equation}
between the $m$ elements at the middle level of the first tube and the $n$ elements at the middle level of the
second, which we picture as two spheres (see Fig.~\ref{fig:spheres}).

\noindent We define $\sigma(\ddot{s}_2/s_2)$ as the standard deviation of $\ddot{s}_2/s_2$ over sprinklings,
$\sigma(\ddot{s}_2/s_2) = (\text{Var}[\ddot{s}_2/s_2])^{1/2}$.

To check whether the detection condition can be met in a sprinkling of $\sim 10^6$ elements, we need
$\sigma(\ddot{s}_2/s_2)$ in terms of $N_{\text{total}}$ and the dimensions of the sprinkling region. Since
$s_2$ is the only quantity in $\ddot{s}_2/s_2$ that fluctuates, we have (Appendix \ref{app:error-propagation})
\begin{equation}
    \sigma\!\left(\ddot{s}_2/s_2\right) \approx \frac{2L}{(\Delta t)^2\,\bar{s}_2^{\,2}}\;\sigma(\bar{d}), \quad \text{where}\quad \bar{s}_2 = \mathbb{E}[\bar{d}].
\end{equation}
The variance of $\bar{d}$ depends on the thickness of the tubes, $\delta$, and on the number of elements, $m$
and $n$, in the two spheres. However, the form of the dependence on $m$ and $n$ depends on what one assumes
about the covariance among the various $d_{ij}$. The detailed calculation in Appendix
\ref{app:error-propagation} shows that
\begin{equation}\label{eq:var_d_bar_main}
    \text{Var}[\bar{d}] = \left(\frac1n+\frac1m\right)\frac13\delta^2 + \mathcal{O}(\delta^4/s^2).
\end{equation}
For $m=n$ this gives 
\begin{equation}
    \sigma(\bar{d}) \approx a_s\,\delta \quad \text{with}\quad a_s \coloneq \sqrt{\frac{2}{3m}}.
\end{equation}

Then, using the continuum values $\left|\ddot{s}_2/s_2\right| = \frac12\omega^2 h_0$ and $\bar{s}_2 \approx
L$, the detection condition becomes a condition on the thickness,
\begin{equation}
    \delta \le \frac{\pi^2}{16}\,\frac{h_0}{a_s}\,L,
\end{equation}
i.e., the thickness must be less than a few times $\frac12 h_0 L$, the change in separation due to the wave.
On the other hand, the thickness depends on $N_{\text{total}}$ and the dimensions of the region through the
power law of Sec.~\ref{sec:thickness}, $\delta = A\,(\pi/24)^{-\beta}\rho^{-\beta}H^{-4\beta+1}$ with $\rho =
N_{\text{total}}/V_{\text{album}}$. We take $V_{\text{album}} = w^2\cross B\cross H$ with $w = a_w\,\delta$
and $B = L + a_w\,\delta$, so that the tubes are just fully contained in the region. Equating the two
expressions for $\delta$, with $A\approx0.3$ and $\beta\approx0.11$ (assuming these values do not change
significantly for a weakly curved spacetime), and writing $L = a_L\lambda$ and $h_0 = a_h\cdot10^{-2}$, gives
\begin{equation}\label{eq:N_total_final}
    \boxed{N_{\text{total}} \sim 7.3 \times 10^{9} \cdot \left(\frac{a_s}{a_h}\right)^7 \cdot a_w^2 \cdot a_L^{-6}.}
\end{equation}
If one takes the most liberal values for these parameters---$a_h=1$ (i.e.\ $h_0=10^{-2}$, far stronger than a
realistic wave's $h_0\sim 10^{-21}$), $a_w=2$, and $a_L=0.5$ (i.e.\ $L=0.5\lambda$, even though $L\ll\lambda$
is required)---this becomes
\begin{equation} \label{eq:N_total_m}
    \boxed{N_{\text{total}} \sim \frac{4.5 \cross 10^{11}}{m^{7/2}}.}
\end{equation}
For $m = 11$, which is larger than the four to five we expect from simulations, this is $N_{\text{total}} \sim
10^8$ elements; for $m\approx 4$--$5$ it is $\sim 10^9$. Both are well beyond the $\sim 4\times10^6$
achievable on a 1\,TB node (Sec.~\ref{sec:limitations}). So even with an ideal distance estimator, end points
known without error, and a wave nineteen orders of magnitude stronger than a realistic one, the method needs
between one and three orders of magnitude more elements than we can sprinkle.

\section{Discussion and outlook}\label{sec:discussion}

Even for these optimistic values of the dimensions and of $h_0$, the $L<\lambda_{\text{GW}}$ regime remains
computationally inaccessible. This means that although the method of finding curvature through tracking the
evolution of the separation between geodesic tubes using a three-point midpoint numerical derivative works in
principle, its efficacy cannot be demonstrated in practice for the case of linearized plane waves. The main
idea behind overcoming the deficiency of the current three-point numerical derivative approach would be to
somehow sample the effects of curvature over a larger volume. One option that can be pursued is to measure
$s(t)$ for all levels of both tubes and then fit a curve to the data points, from which the second derivative
at $t_2$ can be found. Regardless of how the evolution of the separation is measured, one must keep in mind
that the geodesic deviation method inherently requires that the separation between geodesics is small compared
to the curvature length scale of the spacetime.

Moreover, the approach we followed allows us to estimate one component of the Weyl tensor at a time. To
estimate all components at the same time we can think about the interpretation of the Weyl tensor as giving
the evolution of the shear tensor, i.e., the 3D shape of a congruence of geodesics, and track the distances
between four geodesics. The idea then would be to track the six separations of the `tetrahedron' formed by the
four geodesics as it evolves in time. This intuition is also supported by the appropriately titled paper by
Ciufolini and Demianski (1986) \cite{ciufolini1986}, \textit{``How to measure the curvature of space-time''}.

A note on possible distance estimators. One idea would be to modify the causal-overlap method of finding
spatial distances for Minkowski spacetime, introduced by Boguñá and Krioukov \cite{boguna2024}, so that it
could work for Minkowski plus a perturbation. An alternative idea would be to use the shortest return chain
from one point on the tube to anywhere on the neighboring tube, and back anywhere on it. The number of links
on the tube between the departing and arriving elements should be related monotonically to the separation
between the tubes. Thus, if we do this successively for many sections of the first tube, we can get a sense of
change in separation from the change in the size of each section. We believe a similar proposal has been made
by Brightwell and Gregory (1991) \cite{BG}, although theirs is element-to-element, whereas ours is
element-to-chain. One advantage of this approach is that it does not rely on any assumption about the points
on the two geodesics having the same time coordinate. It also would not require the large number of causal
elements in the past required to calculate the causal overlap. On the other hand, it may require the causal
set to be very `tall' in order to accommodate the return signals.

In future work we would like to repeat the process of sprinkling for a class of spacetimes called pp-wave
spacetimes. They too represent waves in spacetime, but instead of the wave being a perturbation on top of
Minkowski spacetime they have the important feature of being \emph{exact} solutions of the Einstein equations
in vacuum. They have the following defining properties (Ref.\ \cite{Stephani2004-tc}, p. 244):
\begin{equation}
    R_{\alpha\beta}=0,\quad k_{\alpha;\nu}=0, \quad k_\alpha k^\alpha=0.
\end{equation}
Here, $k_\alpha$ is a null wave-vector, which in the case of the linear waves of the previous sections was
$(\omega,0,0,\omega)$. The second requirement means that these waves are \textit{plane-fronted waves with
parallel rays (pp-waves).} From Ref.\ \cite{Roche_2023}, ``A plane wave has a planar wavefront (roughly, the
spacelike codimension-2 hypersurface orthogonal to the wave vector), but in general these parallel waves can
have curved wavefronts. Although to obtain wavelike behaviour it is not necessary to demand the wavefront be
flat, it is standard in the field to make this restriction.'' Metrics that satisfy these properties can be
written (locally) in a simplified form (here following the null-coordinate convention of
Ref.\ \cite{Stephani2004-tc}, which differs in sign from Eq.\ \ref{eq:2.1}, where $g_{uv}=-1$) as
\begin{equation}\label{eq:pp_wave_metric}
    \dd s^2=\dd x^2 + \dd y^2+2\,\dd u\,\dd v + H(x,y,u)\,\dd u^2,
\end{equation}
where $H$ must satisfy the condition
$$\left(\frac{\partial^2}{\partial x^2} + \frac{\partial^2}{\partial y^2} \right)H(x,y,u)=0$$ in order for the
metric (\ref{eq:pp_wave_metric}) to satisfy Einstein's equations.

We just need to keep in mind that $H$ necessarily contains singularities \cite{Stephani2004-tc} because no
solution of the condition given above is regular over the whole $(x,y)$ plane; the only exception, $H=H(u)$,
gives flat spacetime. We would then have to use a
sprinkling region that does not contain the singularities and come up with a new way of finding relations
among sprinkled points, to replace the one described in Appendix A.

\appendix
\makeatletter
\renewcommand{\@seccntformat}[1]{%
  \ifnum\pdfstrcmp{#1}{section}=0 Appendix~\csname the#1\endcsname.\quad \else\csname the#1\endcsname\quad\fi}
\makeatother

\section[Appendix A]{Relations between points in a plane gravitational wave spacetime}
\label{app:relations}

\noindent  In this appendix we derive Eq.\ \ref{eq:geodesic-norm-exact} and give a closed form for $I_\pm$. We
use the null coordinates $u=(t-z)/\sqrt2$ and $v=(t+z)/\sqrt2$ throughout, in which the metric is Eq.\
\ref{eq:2.1} with $h_{xx}(u)=h_0\cos(\omega' u)$.

\subsection{The exact interval}\label{app:exact}

Let the geodesic from $p_1$ to $p_2$ be $w^\mu(\lambda) = (u(\lambda), v(\lambda), x(\lambda), y(\lambda))$,
with tangent vector $W^\mu=\dd w^\mu/\dd\lambda$ and the affine parameter normalized so that
$w^\mu(0)=p_1^\mu$ and $w^\mu(1)=p_2^\mu$. We write $g_{xx}(u)\coloneq 1+h_{xx}(u)$ and $g_{yy}(u)\coloneq
1-h_{xx}(u)$ for the two transverse components of the metric.

Along a geodesic we have \cite[Eq.\ 3.167]{carroll_spacetime_2004}
\begin{equation}
    \dd W_\sigma/\dd\lambda=\frac12(\partial_\sigma g_{\alpha\beta})W^\alpha W^\beta,
\end{equation}
so $W_\sigma$ is constant whenever the metric does not depend on $x^\sigma$. The metric does not depend on
$v$, $x$ or $y$. Lowering the indices of $W^\mu$ with the GW metric in Eq.\ \ref{eq:2.1}, the three conserved
components give
\begin{equation}\label{eq:app-conserved}
    W^u=-W_v,\qquad W^x=\frac{W_x}{g_{xx}(u)},\qquad W^y=\frac{W_y}{g_{yy}(u)},
\end{equation}
with $W_v$, $W_x$ and $W_y$ constant. The first says that $u(\lambda)=u_1+\Delta u\,\lambda$, where $\Delta u
= -W_v$. Integrating the second from $\lambda=0$ to $1$, and changing the variable of integration from
$\lambda$ to $u$ with $\dd u=\Delta u\,\dd\lambda$,
\begin{equation}\label{eq:app-Dx}
    \Delta x=x(1)-x(0)=W_x\int_0^1\frac{\dd\lambda}{g_{xx}(u(\lambda))}=\frac{W_x}{\Delta u}\int_{u_1}^{u_2}\frac{\dd u}{g_{xx}(u)}=W_x\,I_+,
    \qquad\text{and similarly}\qquad \Delta y=W_y\,I_-,
\end{equation}
with $I_\pm$ as defined in Eq.\ \ref{eq:geodesic-norm-exact}. For $|h_0|<1$ both $g_{xx}(u)$ and $g_{yy}(u)$
are positive, so $I_\pm>0$ and the two points determine $W_x$ and $W_y$ uniquely. The geodesic between two
points with $\Delta u\neq0$ is therefore unique anywhere in the region covered by the coordinates, which is
requirement (ii) of Sec.~\ref{sec:sprinkling}; for a similar treatment of geodesics in plane waves, see
Ref.~\cite{zhang_soft_2017}.

To find the norm it is convenient to use the covariant components, since three of them are constant. With
$g^{uv}=-1$, $g^{xx}=1/g_{xx}(u)$ and $g^{yy}=1/g_{yy}(u)$,
\begin{equation}\label{eq:app-norm}
    g_{\mu\nu}W^\mu W^\nu=g^{\alpha\beta}W_\alpha W_\beta=-2W^uW^v+\frac{W_x^2}{g_{xx}(u)}+\frac{W_y^2}{g_{yy}(u)} .
\end{equation}
The norm of the tangent vector is constant along the geodesic, so the norm is equal to its integral from
$\lambda=0$ to $1$. So, integrating the individual terms above and summing them also gives us the norm, but
the integration allows us to avoid finding $W^v(\lambda)$ explicitly, which is the only component that is not
constant. Since $W^u=-W_v=\Delta u$ is constant, the first term integrates to $-2\Delta u\int_0^1
W^v\,\dd\lambda=-2\Delta u\,[v(1)-v(0)]=-2\Delta u\,\Delta v$. The other two give $W_x^2I_+$ and $W_y^2I_-$ in
the same way as in Eq.\ \ref{eq:app-Dx}. Eliminating $W_x$ and $W_y$ with Eq.\ \ref{eq:app-Dx},
\begin{equation}
    g_{\mu\nu}W^\mu W^\nu=-2\Delta u\,\Delta v+W_x^2I_++W_y^2I_-=-2\Delta u\,\Delta v+\frac{(\Delta x)^2}{I_+}+\frac{(\Delta y)^2}{I_-},
\end{equation}
which is Eq.\ \ref{eq:geodesic-norm-exact}. Nothing in the argument depends on the particular form of
$g_{xx}(u)$ and $g_{yy}(u)$, so for any metric $-2\,\dd u\,\dd v+g_{xx}(u)\,\dd x^2+g_{yy}(u)\,\dd y^2$ the
same expression holds with $1/I_+$ and $1/I_-$ the harmonic means of $g_{xx}(u)$ and $g_{yy}(u)$ between $u_1$
and $u_2$.

\subsection{Closed form of \texorpdfstring{$I_\pm$}{I+-}}\label{app:Ipm}

The integral in $I_\pm$ can be done in closed form. Writing $\theta=\omega' u$ and $e=\pm h_0$, the
substitution $t=\tan(\theta/2)$ gives, for $|e|<1$,
\begin{equation}\label{eq:app-antider-tan}
    \int\frac{\dd\theta}{1+e\cos\theta}=\frac{2}{r}\arctan\!\Big(k\tan\frac{\theta}{2}\Big),
    \qquad r=\sqrt{1-e^2},\quad k=\sqrt{\frac{1-e}{1+e}}.
\end{equation}
This antiderivative jumps by $2\pi/r$ at every $\theta$ equal to an odd multiple of $\pi$, where
$\tan(\theta/2)$ diverges, so it cannot be used directly when $u_1$ and $u_2$ span several periods. A form
valid for all $\theta$ follows from
$\varphi-\arctan(k\tan\varphi)=\arctan[(1-k)\tan\varphi/(1+k\tan^2\varphi)]$ with $\varphi=\theta/2$. In terms
of $\beta=(1-k)/(1+k)=e/(1+r)$, for $-\pi<\theta<\pi$,
\begin{equation}\label{eq:app-antider}
    \frac{2}{r}\arctan\!\Big(k\tan\frac{\theta}{2}\Big)=\frac{\Phi(\theta)}{r},\qquad
    \Phi(\theta)=\theta-2\arctan\frac{\beta\sin\theta}{1+\beta\cos\theta}.
\end{equation}
Since $|\beta|<1$, the denominator $1+\beta\cos\theta$ never vanishes, so $\Phi$ is smooth for all $\theta$,
and $\Phi'(\theta)=r/(1+e\cos\theta)$ everywhere. Thus $\Phi(\theta)/r$ is an antiderivative on the whole real
line. Using $\dd u=\dd\theta/\omega'$ and $r=\sqrt{1-h_0^2}$ for both signs,
\begin{equation}\label{eq:app-Ipm-closed}
    I_\pm=\frac{\Phi_\pm(\omega' u_2)-\Phi_\pm(\omega' u_1)}{r\,\omega'\,\Delta u},
    \qquad
    \Phi_\pm(\theta)=\theta\mp2\arctan\!\left[\frac{h_0\sin\theta}{1+r\pm h_0\cos\theta}\right],
\end{equation}
which holds however many periods lie between $u_1$ and $u_2$. Because $\Phi_\pm(\theta)-\theta$ is bounded and
periodic, $I_\pm\to1/r$ for both signs when $\omega'\Delta u\gg1$: the difference between the $x$ and $y$
directions averages away. For $h_0\ll1$, $\Phi_\pm(\theta)=\theta\mp h_0\sin\theta+\mathcal{O}(h_0^2)$.

\section[Appendix B]{Error-propagation derivation for $N_{\text{total}}$}
\label{app:error-propagation}
It will be helpful to keep in mind Figs.\ \ref{fig:album-with-spheres} and \ref{fig:spheres} while reading
this appendix.\\
(1) We want to estimate $C_{x0x0}$ at $t_2$ using the equation $\ddot{s}/s = -C_{x0x0}$. As stated above, this
is a gauge-invariant equation in linearized gravity \cite[p.\ 208]{schutz_first_2022}. Here, $\ddot{s}$ and
$s$ are the proper relative acceleration and the proper distance between the geodesics, which are assumed to
be determinable from the ideal element-to-element distance estimator. Later, in steps (13) and (14), to get a
bound on the minimum value for $N_{\text{total}}$ we will use the analytic value $\ddot{s}/s\mid_{t_2} =
-C_{x0x0}\mid_{t_{2}} \approx\frac12\omega^2 h_0$ (at the minimum of the separation, where
$\cos\omega(t-z)=-1$), and also use the analytic value for $s_2 = L(1-\frac12 h_0)$.\\
(2) However, because of the thickness of the tubes at $t_2$, there will be multiple element-to-element
estimates, $d_{ij}$, for $s_2$. If there are $m$ elements in the first sphere and $n$ elements in the second
one, there will be $m \cross n$ distances. Thus we will have an error in the estimate of $\ddot{s}/s$ at
$t_2$.\\
(3) To be more precise, there are two errors: first, the expected fluctuation, Var$[d_{ij}]$, in a single
measurement $d_{ij}$ representing the spread in the values $\{d_{ij}\}$ for a random pair of elements in a
random sprinkling; second, if we take the mean, $\bar{d}$, of the $\{d_{ij}\}$ in a given sprinkling,
\begin{equation}\label{eq:d_bar_def}
    \bar{d} = \frac{1}{mn}\,\sum_i^m \sum_j^n d_{ij},
\end{equation}
where the sums over $i$ ($j$) run over the elements of the first (second) sphere, there will be fluctuations
in the value of $\bar{d}$ as well. We are interested in Var$[\bar{d}]$ (which we expect to be less than
Var$[d_{ij}]$ as derived later) because $\bar{d}$ is our best guess for $s_2$.\\
(4) Thus, for a given sprinkling, to be confident about a detection, we would like
\begin{equation}
    \boxed{\sigma(\ddot{s}_2/s_2) < \ddot{s}_2/s_2}; \quad \sigma(\ddot{s}_2/s_2)\coloneq(\text{Var}[\ddot{s}_2/s_2])^{1/2}, \quad s_2 \coloneq \bar{d}.
\end{equation}
This is our \textbf{detection condition}.\\
(5) We define $\ddot{s}$ using the three-point midpoint formula for the second derivative, i.e.,
\begin{equation}
    \ddot{s}(t_2) = \frac{s(t_1) - 2s(t_2) + s(t_3)}{(\Delta t)^2},
\end{equation}
where $t_1$ and $t_3$ are the times of the bottom and top of the tubes, respectively, $\omega = 2\pi/\lambda$
and $\Delta t = t_3 - t_2 = t_2 - t_1 \equiv \pi/2\omega = \lambda/4$. We choose the height of the box, $H =
t_3 - t_1 = \lambda/2$, and not more because we want to minimize the volume of the sprinkling region while
still getting sufficient change in separation. Moreover, $s(t_1) = s(t_3) = L$ by construction since the tubes
pinch off to individual elements at $t_1$ and $t_3$. We choose the three-point rather than the five-point
formula because we only want there to be one source of error, the estimate of $s_2$.\\
(6) The standard deviation $\sigma(\ddot{s}_2/s_2)$ is ($s_1 = s_3 = L$ and using the properties of the
variance):
\begin{equation}
\sigma\left(\ddot{s}_2/s_2\right) = \left[\frac{1}{(\Delta t)^4}\left(\text{Var}\!\left(\frac{2L}{s_2}\right)\right)\right]^{1/2}.
\end{equation}
Then using the fact that $$\text{Var}\!\left(\frac{1}{X}\right) = \frac{\text{Var}(X)}{\bar{X}^4} +
\mathcal{O}\!\left(\frac{[\text{Var}(X)]^2}{\bar{X}^6}\right);\quad P(X\le 0) \approx 0,$$ we get
\begin{equation*}
    \sigma\left(\ddot{s}_2/s_2\right) = \left[\frac{1}{(\Delta t)^4} \cdot \frac{4L^2}{\bar{s}_2^{\,4}}\,\text{Var}(s_2)\right]^{1/2},
\end{equation*}
that is,
\begin{equation}\label{eq:sigma_ddot_s}
    \boxed{\sigma\left(\ddot{s}_2/s_2\right) = \frac{2L}{(\Delta t)^2 \bar{s}_2^{\,2}}\,\sigma(s_2),}
\end{equation}
where $s_2 = \bar{d}$ and $\bar{s}_2 = \mathbb{E}[\bar{d}]$.\\
(7) Since $s_2 = \bar{d}$, and $\sigma(s_2)=(\text{Var}[s_2])^{1/2}$ we have to find Var$[\bar{d}]$. In this
appendix we will be careful in deriving the variance of $\bar{d}$ because otherwise we might end up with the
wrong dependence of the variance on $m$ and $n$. If one applies Var$[\cdot]$ to the whole expression in Eq.\
\ref{eq:d_bar_def}, then depending on what one assumes about the covariance among the various $d_{ij}$ one can
get, as we will show below, either that:
\begin{equation}
    \text{Var}[\bar{d}]\propto \frac{1}{mn} \frac{\delta^2}{3} \quad \text{or} \quad \text{Var}[\bar{d}]\propto \left(\frac1n + \frac1m\right) \frac{\delta^2}{3}.
\end{equation}
For $m\sim n$ these expressions scale as $1/m^2$ and $1/m$ respectively. Later on we will see that the number
of points, $N_{\text{total}}$, needed to meet the detection condition Eq.\ \ref{eq:detection_condition},
depends on the variance with a power of $\sim 7/2$, thus making a difference of roughly three orders of
magnitude in the value of $N_{\text{total}}$ for $m\sim n\sim 10$ between the two expressions in a realistic
simulation. \\
(8) Instead of dealing with the covariance question directly, we imagine $d_{ij}$ as the norm of the vector
$\vec{d}_{ij}$ defined below.

\noindent There are four vectors in Fig.\ \ref{fig:spheres}: $\vec{a}$ and $\vec{b}$ are the (unknown)
positions of the centers of the spheres respectively, i.e., $s_2 = \mid \vec{b}-\vec{a} \mid \coloneq s$;
$\vec{u}$ and $\vec{v}$ are identical random isotropic vectors centered at the two centers respectively,
representing the random distribution of points, such that $\mathbb{E}[|\vec{v}|] =
\mathbb{E}[|\vec{u}|]=\delta$, and $\mathbb{E}[v_x] = \mathbb{E}[u_x] = 0$ due to isotropy (same for other
components); and we define
\begin{equation}
    \vec{d}_{ij} = (\vec{b}-\vec{a}) + (\vec{v}-\vec{u})\equiv \vec{s}+\vec{\ell},
\end{equation}
where for notational simplicity we have introduced a new vector $\vec{\ell}\coloneq \vec{v}-\vec{u}$,
$|\vec{\ell}|^2 \coloneq \ell^2$, so that $\ell/s \ll 1$ because we assume that $\delta \ll s$ and evidently
$\ell \sim \delta$.

\noindent Now expand the norm as the square root of $\vec{d}_{ij}\cdot \vec{d}_{ij}$ in powers of $(\ell/s)$:
\begin{equation}
    d_{ij} = s\,\left(1 + \frac{\ell_x}{s} + \frac{1}{2s^2}(\ell^2 - \ell_x^2)\right) + \mathcal{O}\left(\frac{\ell^3}{s^2}\right).
\end{equation}
(9) If we take the sum over the $m$ points of the first sphere we get
\begin{equation*}
    \sum_{i}^{m} d_{ij} = m\!\left(s + v_x + \frac{1}{2s}\!\left(v^2 - v_x^2\right)\right) + \sum_{i}\!\left[-u_x + \frac{1}{2s}\!\left(u^2 - 2\,\vec{v}\cdot\vec{u} - u_x^2 + 2v_x u_x\right)\right] + \mathcal{O}\left(\frac{m\ell^3}{s^2}\right).
\end{equation*}
Now if we take the sum over the $n$ points of the second sphere and divide by $mn$
\begin{equation}
\begin{split}\label{eq:d_bar_expansion}
\bar{d} = \frac{1}{mn}\sum_i \sum_j d_{ij}
  &= \frac{\cancel{mn}\cdot s}{\cancel{mn}}
   + \frac{\cancel{m}}{\cancel{m}n}\sum_j\!\left[v_x + \frac{1}{2s}\!\left(v^2 - v_x^2\right)\right] \\
  &\quad + \frac{\cancel{n}}{m\cancel{n}}\sum_i\!\left[-u_x + \frac{1}{2s}\!\left(u^2 - u_x^2\right)\right] \\
  &\quad + \frac{1}{mn\cdot s}\!\left(
      -\sum_j\!(\vec{v}\cdot{\textstyle\sum_i} \vec{u})
      + \sum_j\!(v_x {\textstyle\sum_i} u_x)
    \right) + \mathcal{O}\left(\frac{\ell^3}{s^2}\right).
\end{split}
\end{equation}
(10) Let us take the variance of the whole Eq.\ \ref{eq:d_bar_expansion}. Var$[s]=0$ since $s$ is a constant,
so it drops out. The variance of the last term scales as $1/s^2$ so we consider it of higher order. This
leaves us with the $v$ and $u$ terms. The expressions inside the summations must have the same variances since
the two vectors are i.i.d. The consequence of the summations themselves is to multiply the variance of the
summand by the number of terms in the sum, $n$ for $\sum_j$ and $m$ for $\sum_i$, thus we get
\begin{equation} \label{eq:var_d_bar}
\text{Var}[\bar{d}] = \left(\frac{1}{n} + \frac{1}{m}\right)\text{Var}\!\left[v_x + \frac{1}{2s}\!\left(v^2 - v_x^2\right)\right] + \mathcal{O}\left(\frac{1}{s^2}\right).
\end{equation}
(11) The relationship between $\delta$ and the variances of $\vec{v}$ and $\vec{u}$: \\
\noindent The isotropy of $\vec{v}$ implies that $\mathbb{E}[v_x{}^2] = \mathbb{E}[v_y{}^2] =
\mathbb{E}[v_z{}^2]$, thus $\mathbb{E}[v_x{}^2] = \frac13\mathbb{E}[|\vec{v}|^2]$. Let $\text{Var}[|\vec{v}|]
\equiv \sigma_r{}^2 = \mathbb{E}[|\vec{v}|^2] - (\mathbb{E}[|\vec{v}|])^2$, then
\begin{equation}
\delta \equiv \mathbb{E}[|\vec{v}|] \implies \mathbb{E}[v_x{}^2] = \frac13\,\mathbb{E}[|\vec{v}|^2] = \frac13\,(\sigma_r{}^2 + \delta^2).
\end{equation}
Finally, since Var$[v_x] = \mathbb{E}[v_x^2] - \mathbb{E}[v_x]^2$ and $\mathbb{E}[v_x] = 0$
\begin{equation}\label{eq:var_v_x}
    \boxed{\text{Var}[v_x] = \frac13\,(\sigma_r{}^2 + \delta^2) \gtrsim \frac13\delta^2} \quad \text{if } \sigma_r{}^2 < \delta^2.
\end{equation}
Also, since $\vec{\ell}=\vec{v}-\vec{u}$, where $\vec{v}$ and $\vec{u}$ are i.i.d., and $\ell^2 = v^2 + u^2 -
2\vec{v}\cdot\vec{u}$ and the expectation value of the last term is zero, $\ell^2$ is of the order of
$\delta^2$. \\
(12) The variance term in Eq.\ \ref{eq:var_d_bar},
\begin{equation}
    \text{Var}\left[v_x + \frac{1}{2s}\!\left(v^2 - v_x^2\right)\right]= \text{Var}\left[-u_x + \frac{1}{2s}\!\left(u^2
    - u_x^2\right)\right],
\end{equation}
is of the form Var$[A+B]=\text{Var}[A]+\text{Var}[B]+2\,\text{Cov}[A, B]$, where Cov$[A,
B]=\mathbb{E}[AB]-\mathbb{E}[A]\mathbb{E}[B]$. But both
\begin{equation}
    \mathbb{E}[v_x(v^2-v_x^2)] = 0 \quad \text{and} \quad \mathbb{E}[v_x]\mathbb{E}[(v^2-v_x^2)]=0
\end{equation}
because of the isotropy of $v_x$ (it is as likely to be positive as negative), thus the covariance is zero and
\begin{equation}
    \text{Var}\left[v_x + \frac{1}{2s}\!\left(v^2 - v_x^2\right)\right] = \text{Var}[v_x] + \frac{1}{4s^2}\text{Var}[v^2-v_x^2].
\end{equation}
Ignoring terms of $\mathcal{O}(1/s^2)$, finally we get
\begin{align}
    \text{Var}[\bar{d}] &= \left(\frac1n+\frac1m\right)\text{Var}[v_x] + \mathcal{O}(1/s^2)\\ \notag
    \Aboxed{\text{Var}[\bar{d}]&= \left(\frac1n+\frac1m\right)\frac13\delta^2 + \mathcal{O}(\delta^4/s^2)} \quad \text{using Eq. \ref{eq:var_v_x}}.
\end{align}
We have also confirmed this result by running numerical simulations of the ``two-spheres problem''. The code
can be found at the \texttt{GitHub} repository \url{https://github.com/sniafrmpk/cst}.\\
\noindent (13) Putting it all together, we have using Eqs.\ \ref{eq:var_d_bar}, \ref{eq:sigma_ddot_s},
\begin{equation}
\sigma\left(\ddot{s}_2/s_2\right) \approx \frac{2L}{(\Delta t)^2 \bar{s}_2^{\,2}}\,\sqrt{\frac{m+n}{3mn}}\,\delta + \mathcal{O}\left(\frac{1}{s^4}\right).
\end{equation}
Recall, $\bar{s}_2 \equiv \mathbb{E}[\bar{d}]$. One can use Eq.\ \ref{eq:d_bar_expansion} to find
$\mathbb{E}[\bar{d}]$ to get
\begin{equation}
    \frac{1}{\bar{s_2}^2} = \frac{1}{\mathbb{E}[\bar{d}]^2} = \frac{1}{s^2}\left[1 - \frac{4\delta^2}{3s^2} +\mathcal{O}\left(\frac{1}{s^4}\right)\right].
\end{equation}
Taking $s= L\!\left(1 - \frac12 h_0\right)$, the analytic value of the separation at $t_2$, $h_0 \approx
\delta^2/L^2 \ll \delta/L$, and $\Delta t = \lambda/4$, we get
\begin{align}
\sigma\left(\ddot{s}_2/s_2\right) &\approx \frac{32}{L\,\lambda^2}\,\sqrt{\frac{m+n}{3mn}}\,\delta + \mathcal{O}\left(h_0\sim \frac{\delta^2}{L^2}\right), \notag \\
\Aboxed{\sigma\left(\ddot{s}_2/s_2\right) &\approx \frac{32}{L\,\lambda^2}\,\sqrt{\frac{2}{3m}}\,\delta + \mathcal{O}\left(h_0\sim \frac{\delta^2}{L^2}\right)} \quad \text{for} \quad m=n.
\end{align}
(14) Recall, we want $\sigma\left(\ddot{s}_2/s_2\right) < \ddot{s}_2/s_2$ to be able to detect the curvature.
If we take
$$\ddot{s}_2/s_2 \approx \frac12 \omega^2 h_0,$$ i.e., the analytic value and use the expression for
$\sigma\left(\ddot{s}_2/s_2\right)$ given above, then we get the condition\footnote{$a_s$ would have been
proportional to $1/m$ if the covariances between the $d_{ij}$'s had been zero, which, as shown in step (7),
would have meant a difference of three orders of magnitude in the value of $N_{\text{total}}$.}
\begin{equation}\label{eq:condition_delta}
    \boxed{\delta \le \frac{L\pi^2}{16}\frac{h_0}{a_s},} \quad \text{where } a_s = \sqrt{\frac{2}{3m}}.
\end{equation}
(15) Recall also from Sec.~\ref{sec:thickness}, that
\begin{equation}\label{eq:thickness_trend}
    \delta = A \left(\frac{\pi}{24}\right)^{-\beta} \rho^{-\beta} H^{-4\beta + 1}.
\end{equation}
But since $\rho = N_{\text{total}}/V_{\text{album}}$, where $V_{\text{album}}$ is the volume of the
album-shaped sprinkling region, we can write $\delta$ in terms of $N_{\text{total}}$ and the dimensions of the
album.\\
(16) Let $V_{\text{album}} = w^2\cross B\cross H$ and
\begin{equation}
    w = a_w\,\delta, \quad B = L + w = L + a_w\,\delta; \quad a_w \sim 2.
\end{equation}
We choose to define things this way so that the tubes are as far away as possible from each other and they are
just fully contained in the region; the full intervals formed by the base and top points of the tubes will not
be included in the sprinkling region because the tubes form independently of the region surrounding them.
Then, Eq.\ \ref{eq:thickness_trend} becomes
\begin{equation}\label{eq:delta_in_terms_of_dimensions}
    \boxed{\delta = A \left(\frac{N_{\text{total}}\,\pi}{24\cdot a_w{}^2\delta^2\cdot(L+a_w\delta)}\right)^{-\beta}\, H^{-3\beta + 1}.}
\end{equation}
(17) So, we have two equations, \ref{eq:condition_delta} and \ref{eq:delta_in_terms_of_dimensions}, for how
$\delta$ must be so that it obeys the numerical trend found in Sec.~\ref{sec:thickness} and so that it is
small enough to allow a detection of the deviation between the tubes due to the gravitational wave. They can
be equated to get the minimum value of $N_{\text{total}}$ that allows $\delta$ to satisfy both conditions. If
we introduce more constants to determine the sensitivity of $N_{\text{total}}$ to the various dimensions and
parameters, specifically,
\begin{equation}
    L = a_L \lambda, \quad h_0 = a_h\cdot 10^{-2},
\end{equation}
and use the fact that numerically we have found
\begin{equation}
    A \sim 0.3, \quad \beta \sim 0.11,
\end{equation}
for the case of 4D Minkowski spacetime, and that these values are not expected to change significantly for a
weakly curved spacetime, we can solve for $N_{\text{total}}$ to get (omitting the algebraic details):
\begin{equation}\label{eq:N_total_var}
    \boxed{N_{\text{total}} \sim 7.3 \cross 10^{9} \cdot \left(\frac{a_s}{a_h}\right)^7 \cdot a_w^2 \cdot a_L^{-6}.}
\end{equation}
This is Eq.\ \ref{eq:N_total_final} of Sec.~\ref{sec:feasibility}. In the main text and in this appendix, we
make optimistic assumptions about the values of the dimensionless constants $a_s$, $a_h$, $a_w$, and $a_L$ and
about the variance of $\bar{d}$, and we take two of the three distances in the three-point derivative for
estimating $\ddot{s}_2$, namely $s_1$ and $s_3$, to be known exactly. Even so, we find the value of
$N_{\text{total}}$ to be inaccessibly large.

\section*{Acknowledgements}

We would like to thank Dr.\ Leo Stein for insightful suggestions and discussions. This work was conducted as
part of N.I.'s doctoral studies, and N.I. would like to thank the Department of Physics and Astronomy and the
Graduate School at the University of Mississippi for their support. We also acknowledge the support of
generative AI, especially ChatGPT 4.1, Claude Sonnet, Opus and Fable models.
\bibliographystyle{unsrt_spaced}
\bibliography{bibliography}

\end{document}